\nonstopmode
\documentclass[12pt]{article}

\usepackage{amsmath}
\usepackage{amsfonts}
\usepackage{amssymb}
\usepackage{bbm}
\usepackage{a4wide}
\usepackage{graphicx}
\graphicspath{{./}{../graphics/}}
\usepackage[margin=10pt,font=small]{caption}
\usepackage{hyperref}

\newcommand{\R}{\mathbb{R}}

\newcommand{\Z}{\mathbb{Z}}
\newcommand{\dif}{\,{\rm d}}
\DeclareMathOperator{\vol}{vol}
\DeclareMathOperator{\real}{Re}

\newcommand{\arXiv}[2]{\href{http://arxiv.org/pdf/#1}
  {\texttt{arXiv:#1 #2}}}

\newcommand{\image}[2]{\includegraphics[#1]{#2}}

\newcommand{\figI}{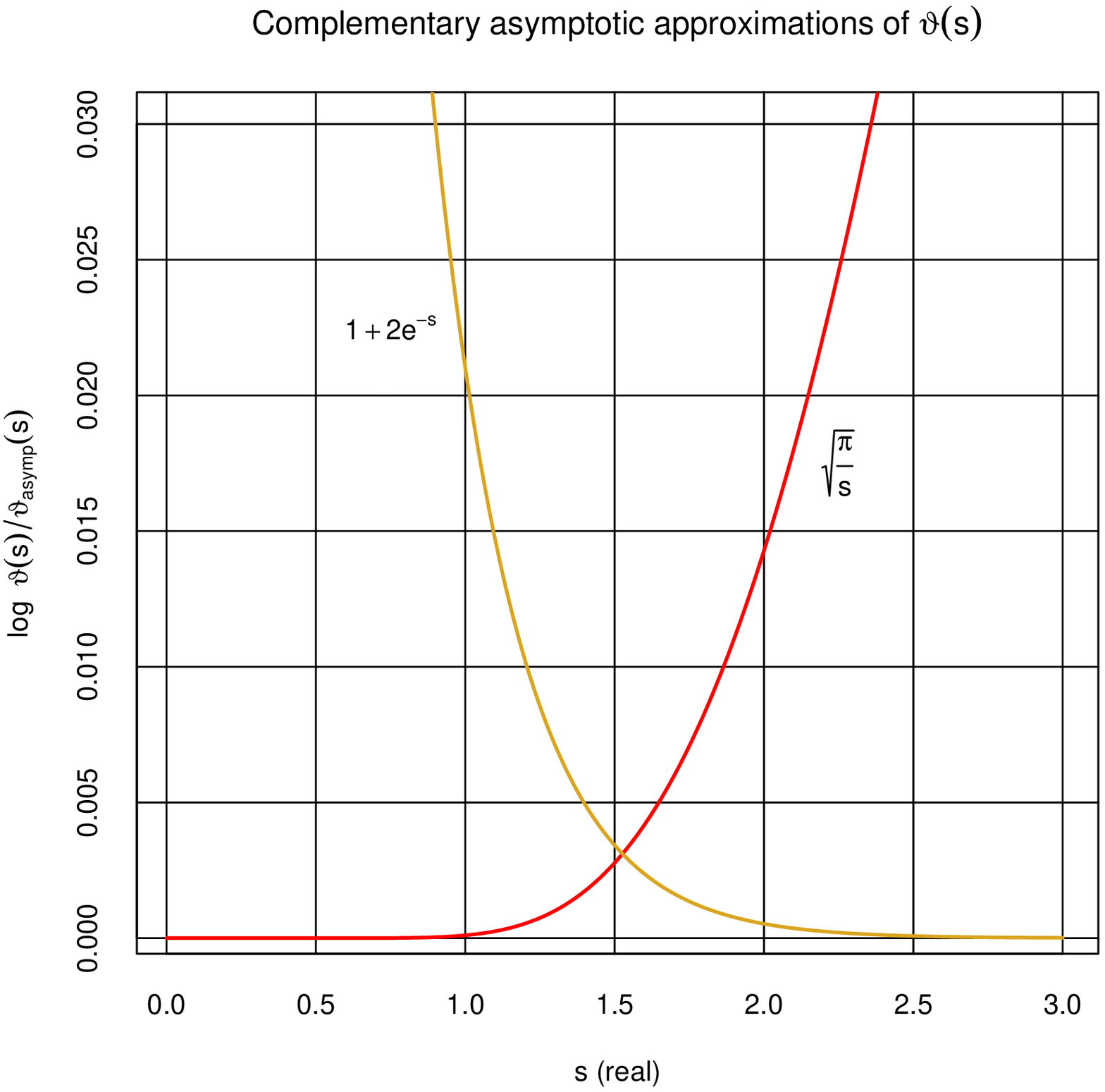}
\newcommand{\figII}{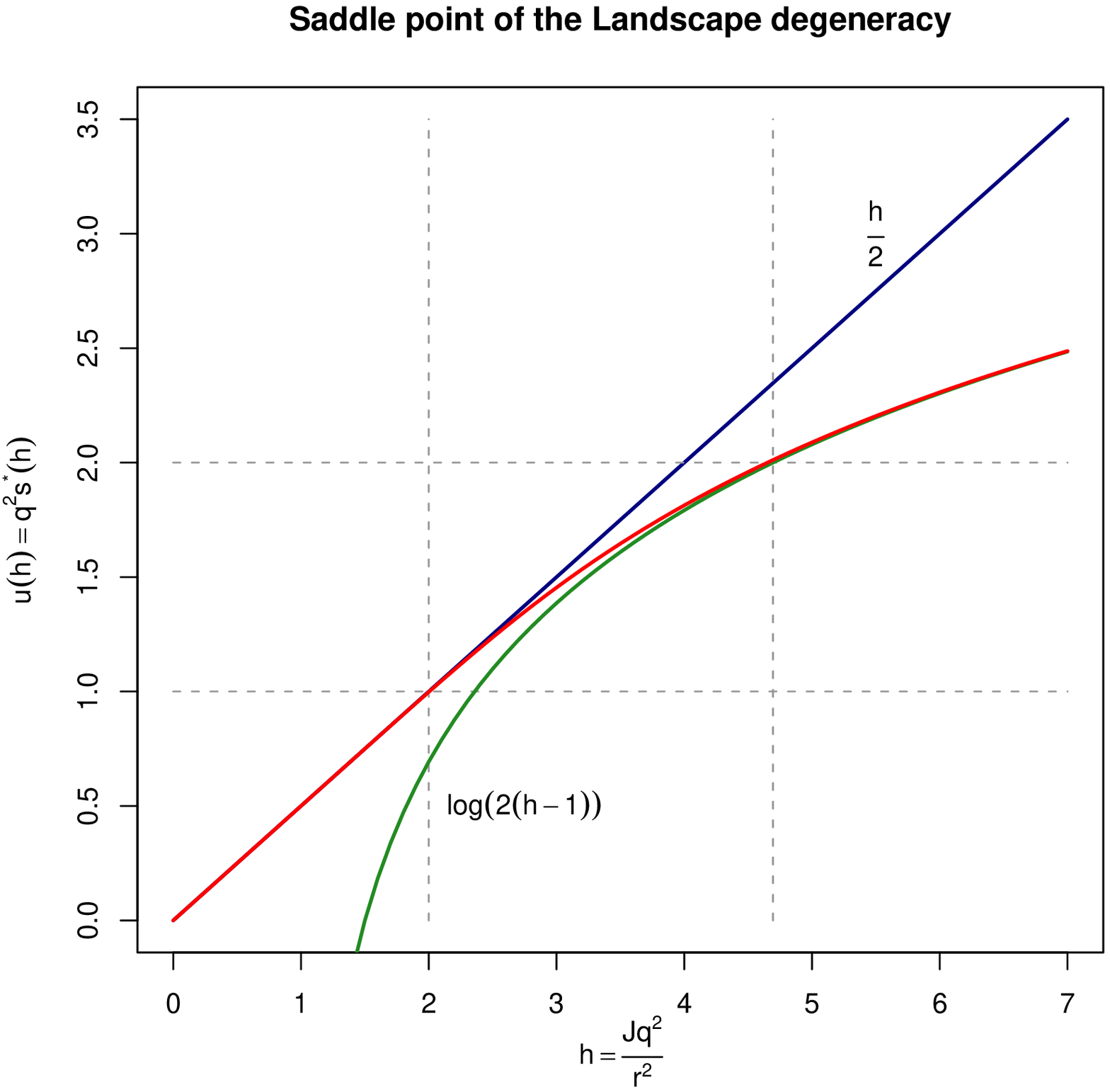}
\newcommand{\figIII}{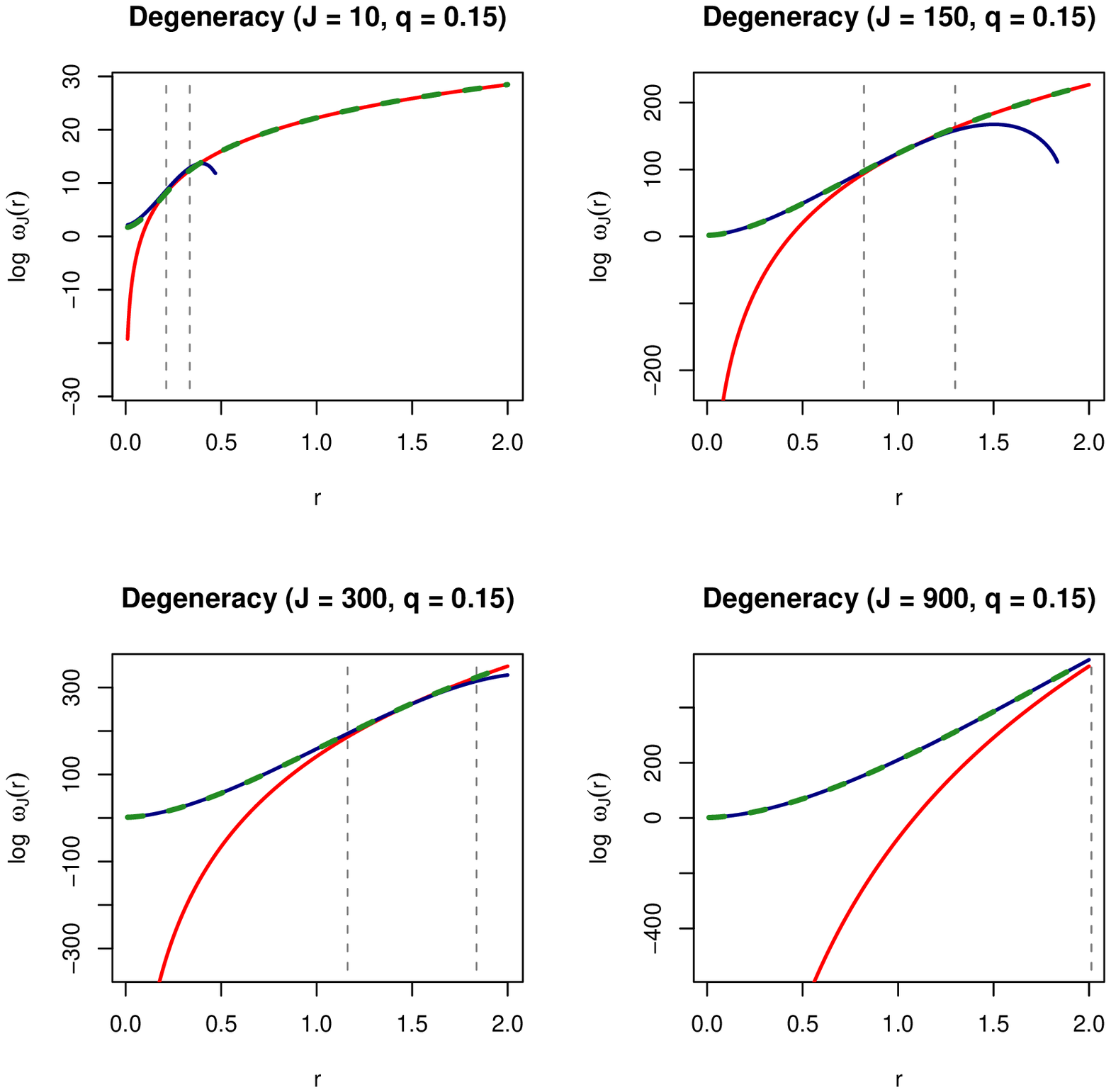}
\newcommand{\figIV}{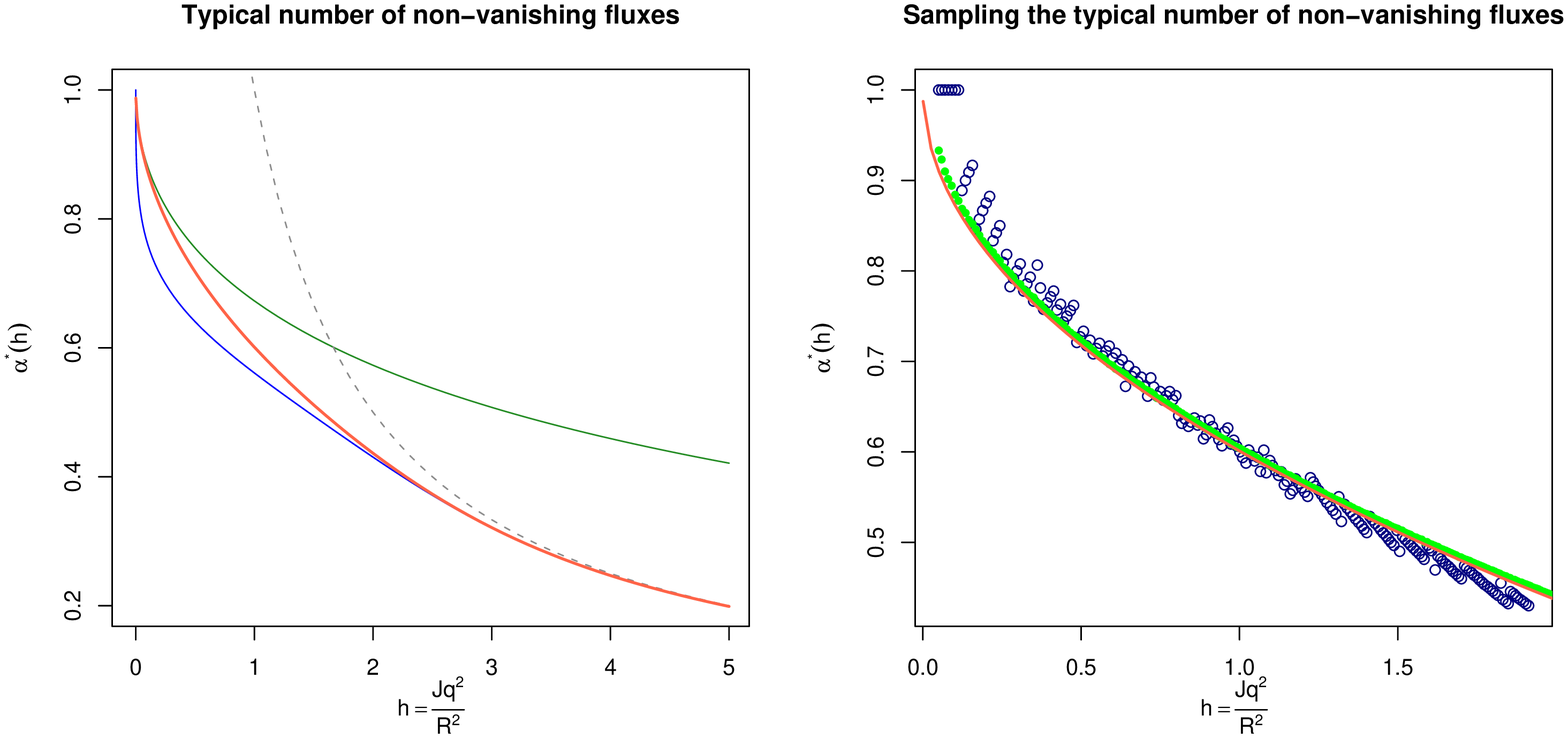}
\newcommand{\figV}{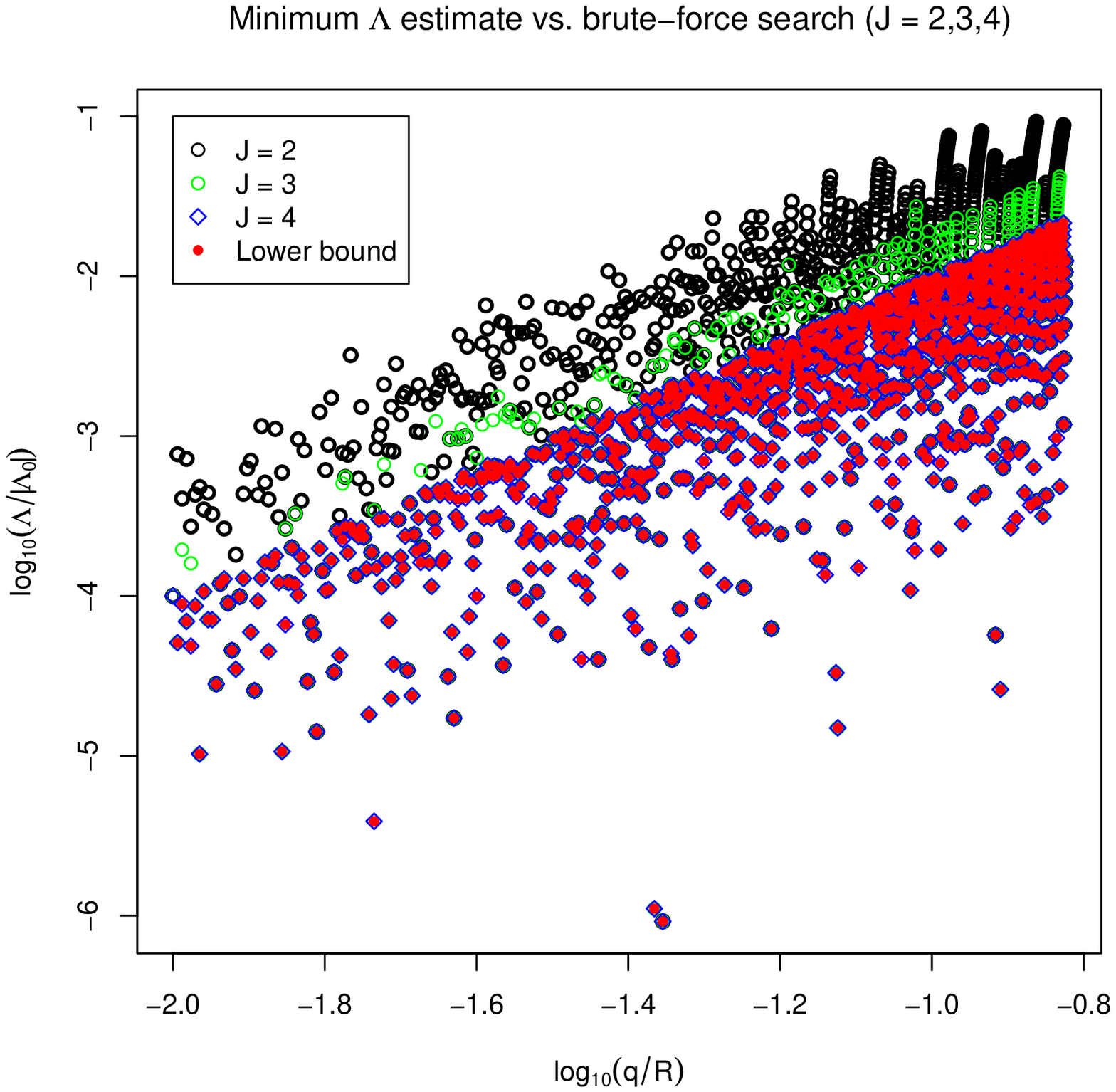}
\newcommand{\figVI}{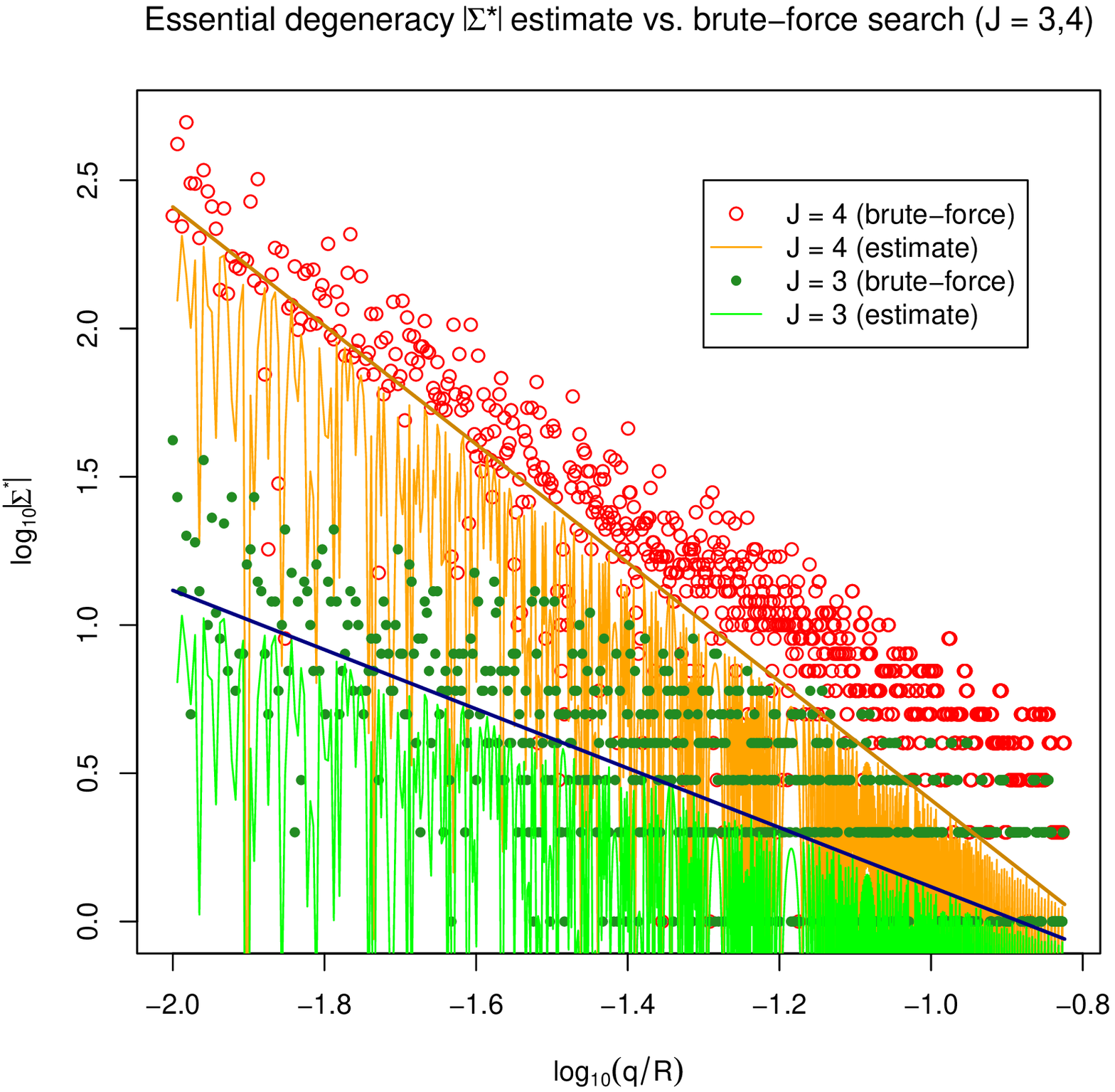}
\newcommand{\figVII}{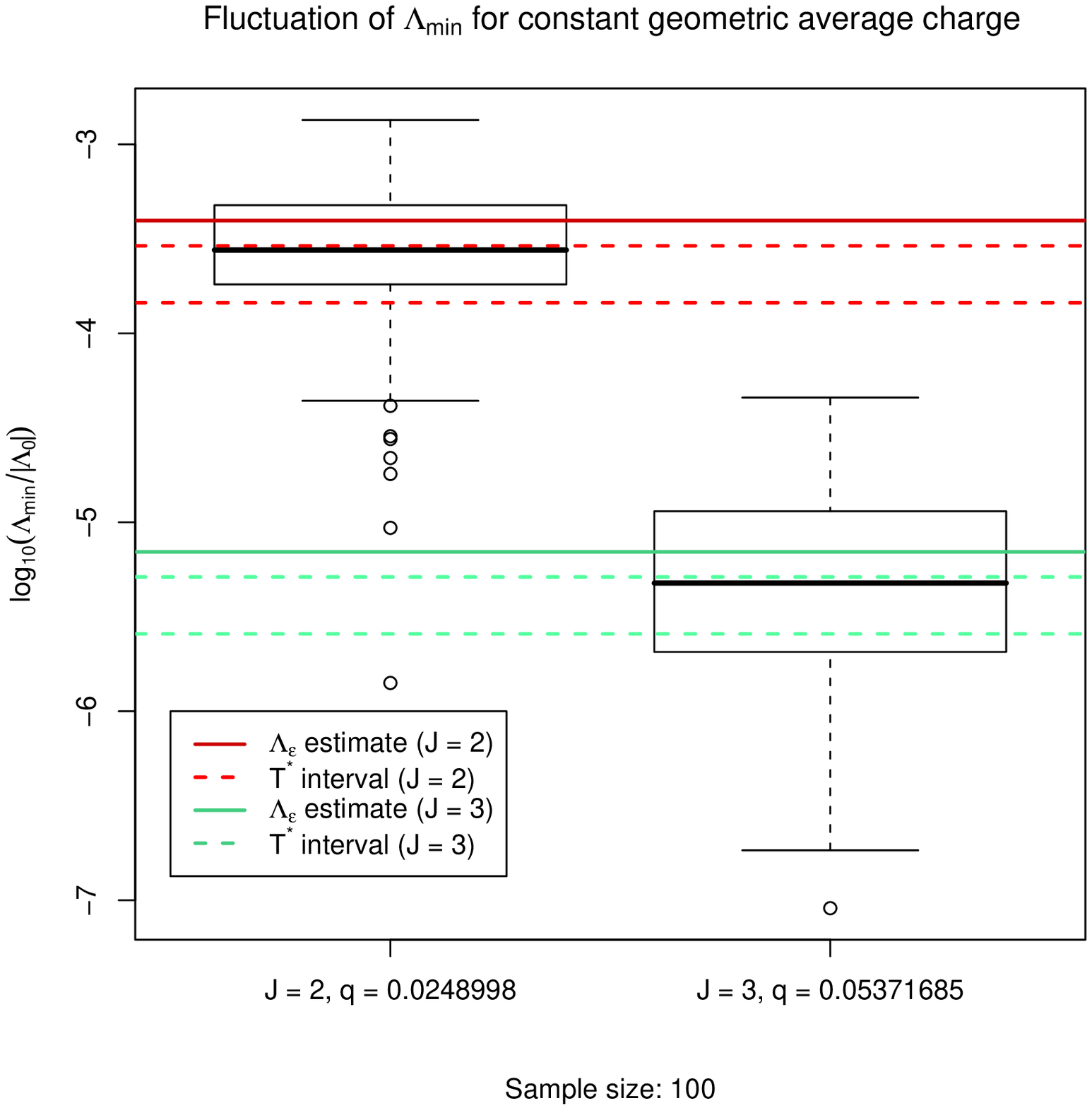}
\newcommand{\figVIII}{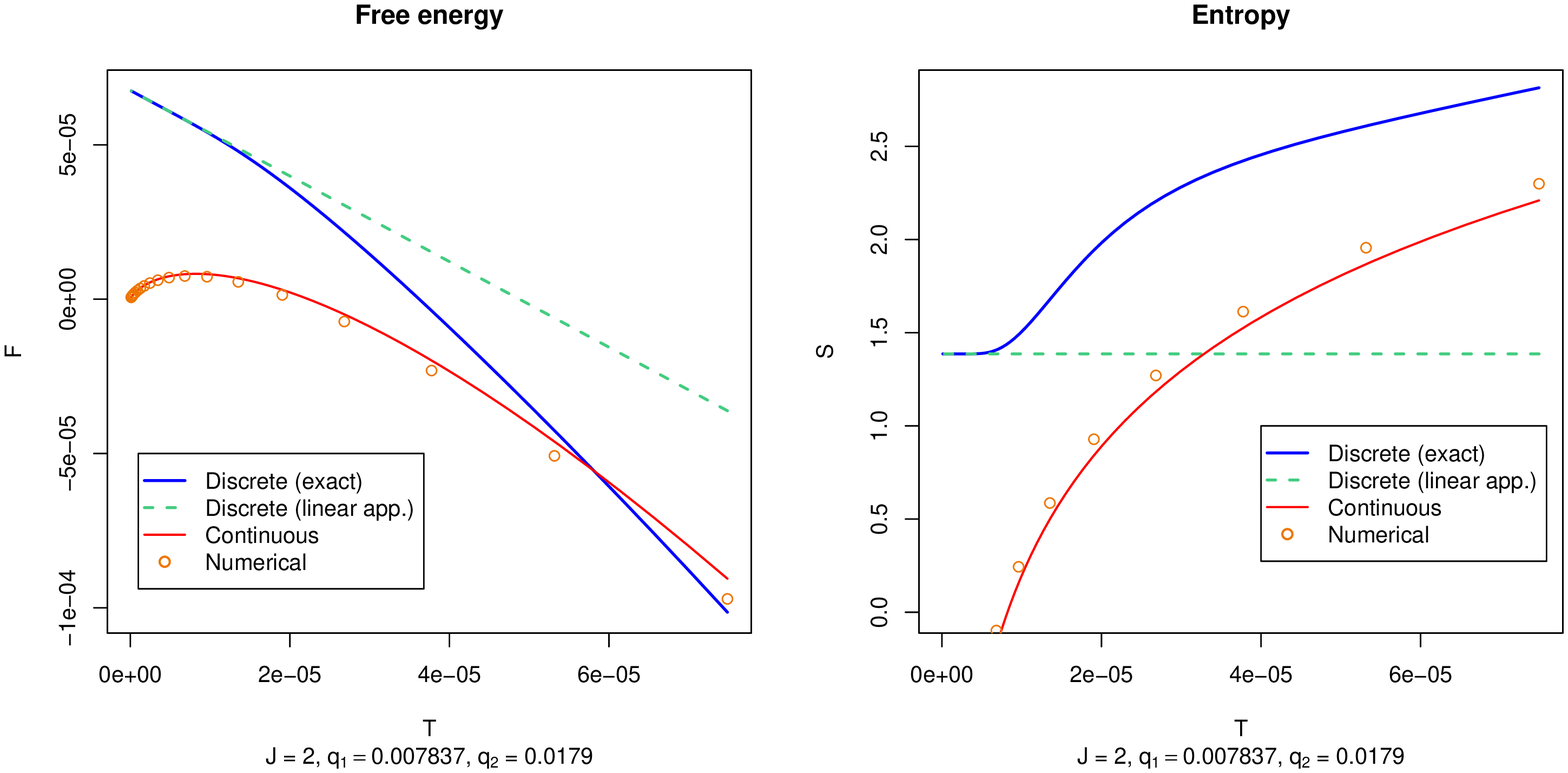}
\newcommand{\figIX}{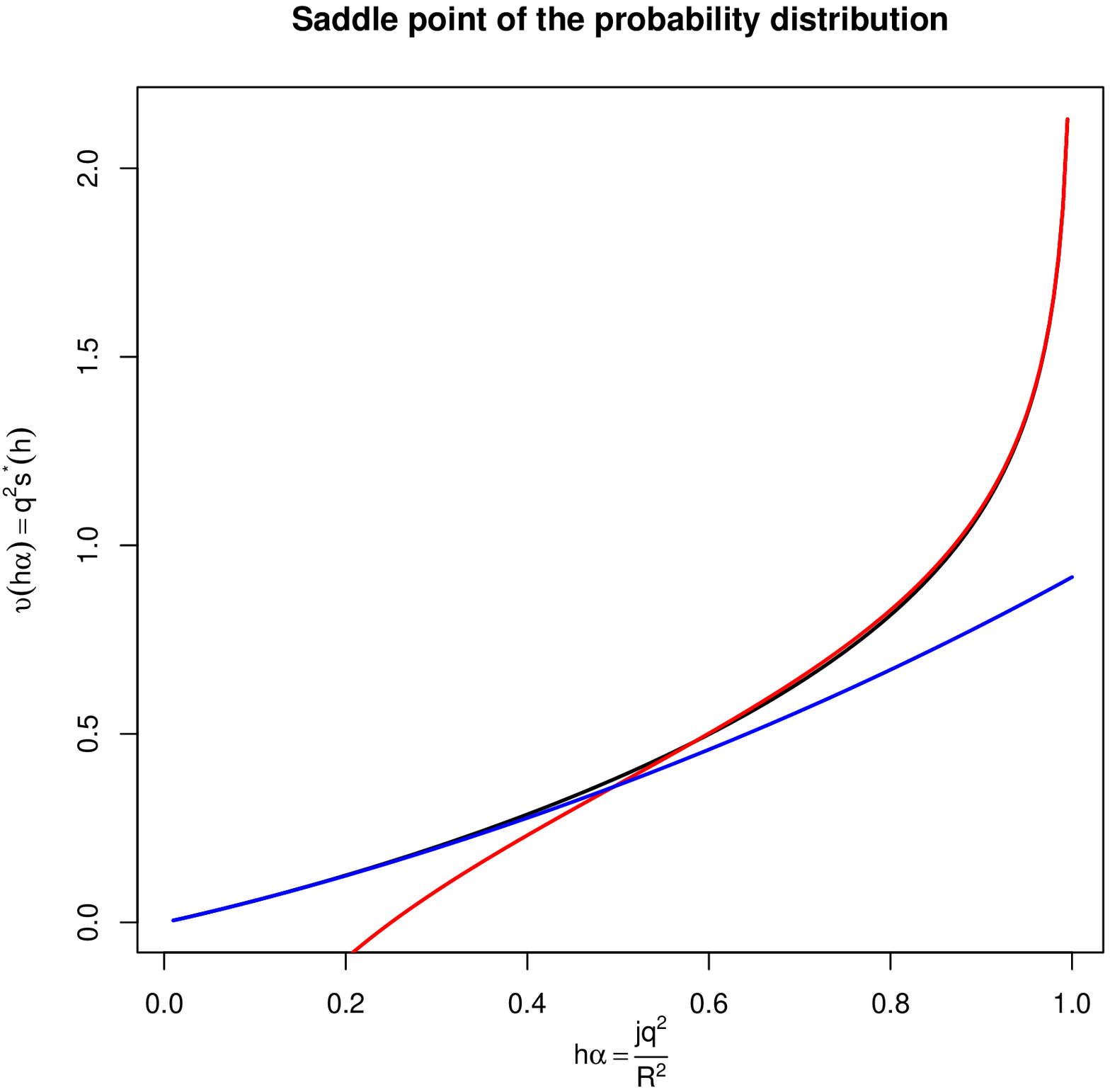}
\newcommand{\figX}{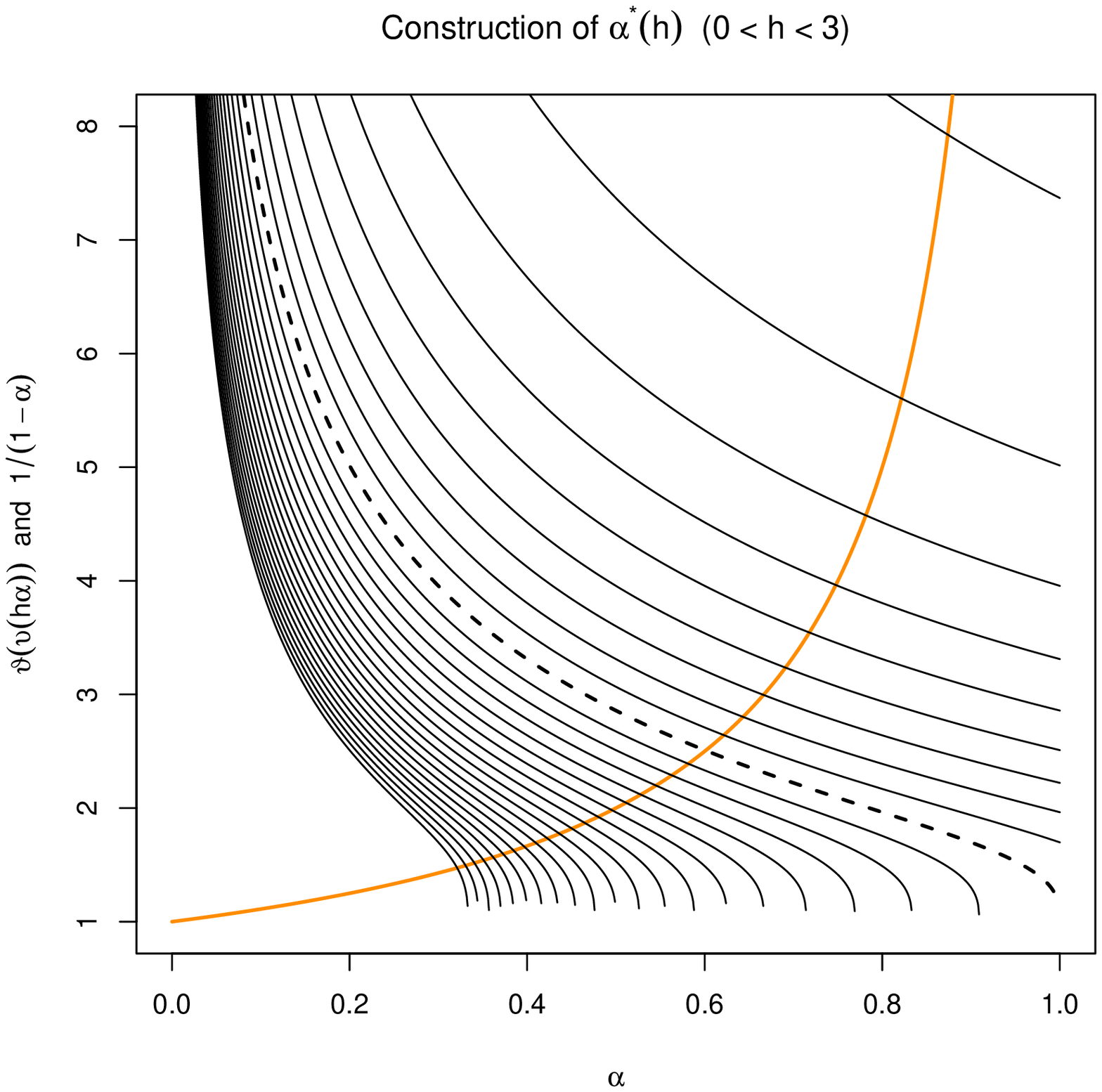}
\newcommand{\figXI}{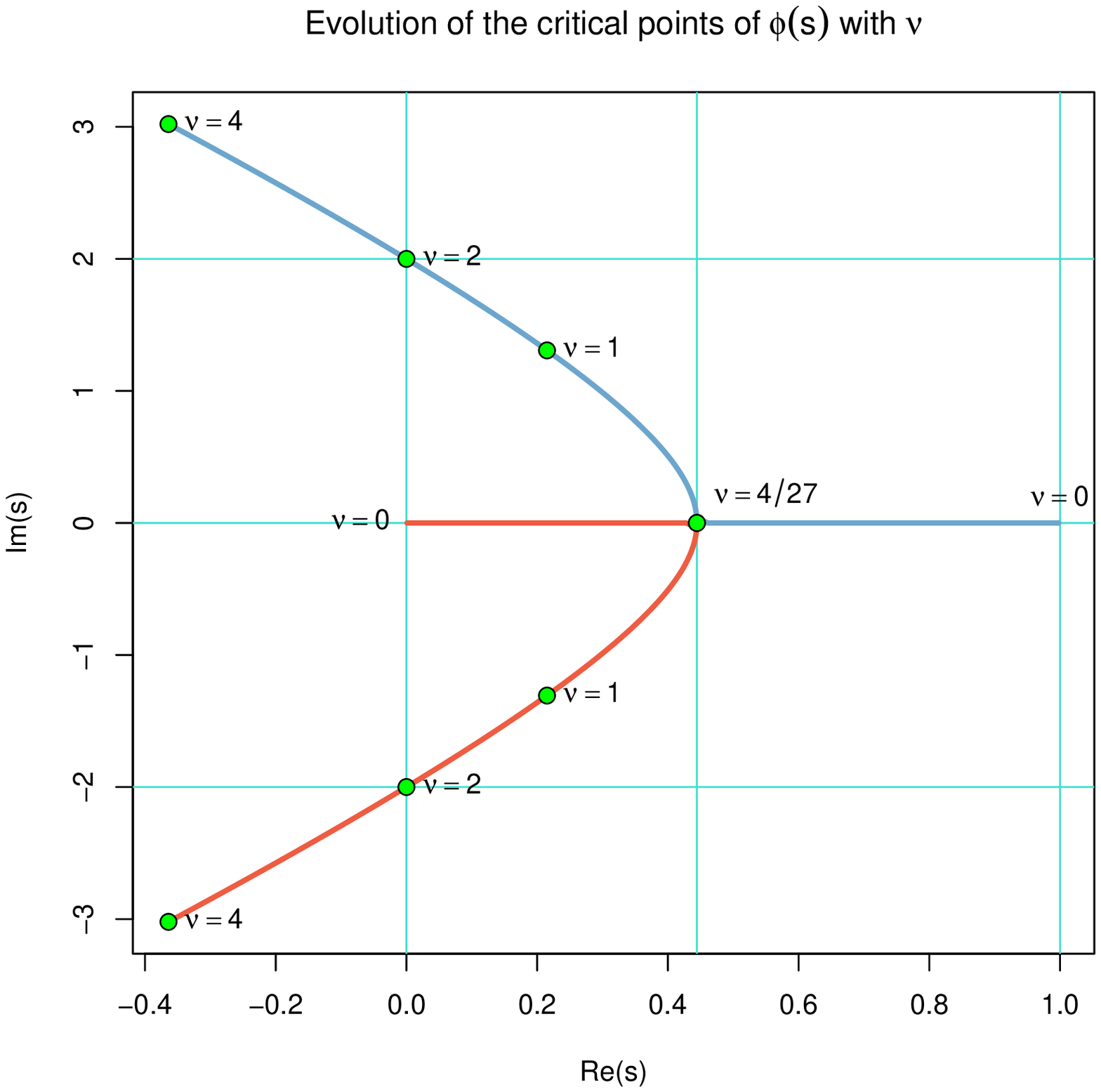}

\title{%
  Applications of an exact counting formula in the Bousso-Polchinski Landscape%
}

\author{%
  C\'esar Asensio\thanks{\texttt{casencha}\textbf{@}\texttt{unizar.es}}%
  \qquad and \qquad %
  Antonio Segu\'{\i}\thanks{\texttt{segui}\textbf{@}\texttt{unizar.es}}%
  \\
  \\
  \emph{Departamento de F\'{\i}sica Te\'orica, Universidad de Zaragoza}%
}

\begin{document}
\maketitle
\begin{abstract}
  The Bousso-Polchinski (BP) Landscape is a proposal for solving the
  Cosmological Constant Problem.  The solution requires counting the
  states in a very thin shell in flux space.  We find an exact formula
  for this counting problem which has two simple asymptotic regimes,
  one of them being the method of counting low $\Lambda$ states given
  originally by Bousso and Polchinski.  We finally give some
  applications of the extended formula: a robust property of the
  Landscape which can be identified with an effective occupation
  number, an estimator for the minimum cosmological constant and a
  possible influence on the KKLT stabilization mechanism.
\end{abstract}

\section{Introduction}
\label{sec:intro}

The eternal inflation picture of the multiverse consists of de Sitter
bubbles nucleating in certain vacuum state of very high energy density
\cite{EtInf-1,EtInf-2,EtInf-3,EtInf-4,EtInf-5}.  Bubbles can be
created inside other bubbles, and this provides a dynamical relaxation
mechanism which gives rise to an average neutralization of the
cosmological constant \cite{BT-1,BT-2}.  We may wonder if it's
possible to formulate a model of eternal inflation with relaxation
containing vacuum states of a cosmological constant as small as the
observed value\footnote{We use reduced Planck units in which $8\pi G =
  \hbar = c = 1$.}  \cite{SN-1,SN-2}
\begin{equation}
  \label{eq:1}
  \Lambda_{\text{obs}} = 1.5\times10^{-123}\,,
\end{equation}
without fine-tuning the parameters of the model and avoiding the need
of invoking the anthropic principle \cite{BaTi,WW,Anthr}, which has
been used to explain why we are living in such a special region of the
multiverse.  The smallness of the number in (\ref{eq:1}) is the
\emph{cosmological constant problem} \cite{WW2,B-CC}.  An attempt for
a solution is given in the Bousso-Polchinski Landscape \cite{BP}, in
which a large amount $J$ of quantized fluxes of charges
$\{q_j\}_{i=1,\cdots,J}$ leads to an effective cosmological constant
\begin{equation}
  \label{eq:2}
  \Lambda = \Lambda_0 + \frac{1}{2}\sum^J_{j=1}n_j^2 q_j^2\,.
\end{equation}
In (\ref{eq:2}), $\Lambda_0$ is a negative number of order $-1$, and
the integer $J$-tuple $(n_1,\cdots,n_J)$ characterizes each of the
vacua of the Landscape.  Without fine-tuning, for large $J$ and
incommensurate charges $\{q_j\}$ this model contains states of small
$\Lambda$.  The problem arises now as how to count them.
Unfortunately, the amount of these \emph{anthropic} states is expected
to be very small as compared to the total number of vacua in the
Landscape \cite{BY}, and therefore it seems not to be other way out
but invoking the anthropic principle to explain the value
(\ref{eq:1}).

The states in the Bousso-Polchinski Landscape can be viewed as nodes
of a lattice in flux space $\R^J$.  We call this lattice $\mathcal{L}$
and the charges $q_i$ are the periods of $\mathcal{L}$, that is,
\begin{equation}
  \label{eq:3}
  \mathcal{L} = \bigl\{(n_1q_1,\cdots,n_Jq_J)\in\R^J\colon
    n_1,\cdots,n_J\in\Z\bigr\}\,.
\end{equation}
A single state $\lambda$ in this lattice is characterized by the
quantum numbers $(n_1,\cdots,n_J)$, and its cosmological constant is,
according (\ref{eq:2}),
\begin{equation}
  \label{eq:4}
  \Lambda(\lambda) = \Lambda_0 + \frac{1}{2}\|\lambda\|^2\,.
\end{equation}
Vacuum states in the Bousso-Polchinski Landscape are defined in the
semiclassical approximation as stationary points of an effective
action.  If we consider two neighbor states of very high $\Lambda$
(neighbor states have quantum numbers $\{n_j\}$ which are different at
one place and by one unit) we find that the energy barrier separating
them is small, that is, the mediating Brown-Teitelboim instanton has a
comparatively low action.  These states are not isolated and
consequently the semiclassical approximation breaks down for them.
Moreover, when $\Lambda$ reaches a value near the Planck energy
density (of order $\Lambda\approx1$), quantum gravity effects become
important, and some approximations made in the Bousso-Polchinski model
(as neglecting the backreaction effect, for example) are no longer
valid.

Thus the Bousso-Polchinski Landscape is a finite subset (yet an
enormous one, a commonly quoted number being $10^{500}$
\cite{Count-1,Count-2,Count-3,Count-4}) of the lattice (\ref{eq:3}),
comprising the nodes with cosmological constant smaller than some
value $\Lambda_1=\mathcal{O}(1)$.

We will review the counting argument of Bousso and Polchinski
\cite{BP,Shenker}.  Around each node $\lambda$ of the lattice
$\mathcal{L}$ we have a Voronoi cell which is a translate of the
parallelotope $Q=\prod^J_{i=1}[-\frac{q_i}{2},\frac{q_i}{2}]$ of
volume $\vol Q=\prod^J_{i=1}q_i$.  On the other hand, each value of
the cosmological constant $\Lambda_0 \le \Lambda \le \Lambda_1$
defines a ball $\mathcal{B}^J(R_\Lambda)$ in flux space of radius
$R_\Lambda=\sqrt{2(\Lambda-\Lambda_0)}$, and whose volume is
\begin{equation}
  \label{eq:5}
  \vol \mathcal{B}^J(R_\Lambda) = \frac{R_\Lambda^J}{J}\vol S^{J-1}\,,
\end{equation}
where the volume of the $J-1$ sphere is
\begin{equation}
  \label{eq:6}
  \vol S^{J-1} =
  \frac{2\pi^{\frac{J}{2}}}{\Gamma\bigl(\frac{J}{2}\bigr)}
  \,.
\end{equation}
The BP counting argument consists of computing the number of states
inside a ball of any radius $r$, which will be called $\Omega_J(r)$,
by taking the quotient between the volume of the ball and the volume
of the cell of a single state:
\begin{equation}
  \label{eq:7}
  \Omega_J(r) = \frac{\vol\mathcal{B}^J(r)}{\vol Q}\,.
\end{equation}
Plugging in some numbers, if $\Lambda_0=-1$, $\Lambda_1=1$, $J=300$
and $q_i=\frac{1}{10}$, we have a BP Landscape with
\begin{equation}
  \label{eq:8}
  \Omega_{300}(2) = 1.3\times10^{202}\quad\text{states}\,.
\end{equation}
This is certainly very small when compared with $10^{500}$.  But if we
take a charge ten times smaller $q_i=0.01$, we obtain $\Omega_{300}(2)
= 1.3\times10^{502}$.  Nevertheless, there is another argument which
gives a more impressive number.  The quotient between $2R_{\Lambda_1}$
and $q_i$ gives the number of nodes that fit in axis $i$.  So, the
total number of nodes is the product of all these quotients, which is
also the volume of the hypercube of side $2R_{\Lambda_1}$ divided by
the volume of the cell.  With the same numbers this quantity is
$4.1\times10^{480}$.  This argument is wrong though, because the vast
majority of the nodes of the hypercube lie outside the sphere of
radius $R_{\Lambda_1}$, and thus they are not states of the Landscape.

We may wonder how many states there are with values of $\Lambda$
comprised between 0 and a small positive value of the cosmological
constant $\Lambda_\varepsilon$.  Following the same argument as in the
preceding paragraph, we compute the quotient between the volume of the
shell comprised between radii $R_0 = \sqrt{2|\Lambda_0|}$ and
$R_\varepsilon = \sqrt{2(\Lambda_\varepsilon - \Lambda_0)} \approx R_0
+ \frac{\Lambda_\varepsilon}{R_0}$ and we obtain
\begin{equation}
  \label{eq:9}
  \mathcal{N}_\varepsilon = \Omega_J(R_\varepsilon) - \Omega_J(R_0)
  \approx R_0^{J-2}\,\frac{\vol S^{J-1}}{\vol Q}\,\Lambda_\varepsilon
  \,.
\end{equation}
If $\Lambda_\varepsilon$ is of the order of the observed value
(\ref{eq:1}), with the previous numbers we obtain
$\mathcal{N}_\varepsilon = 2.1\times10^{36}$.  Thus, if the degeneracy
of these states is smaller than this number, we can find states with a
realistic cosmological constant.

Nevertheless, as the authors of \cite{BP} point out, this argument is
not valid when any of the charges $q_i$ exceed $R_0/\sqrt{J}$.  We can
see that strange things happen as $J$ grows with all the charges
fixed.  There is a critical value of $J$ above which the volume of the
sphere is smaller than the volume of a single cell,
\begin{equation}
  \label{eq:10}
  \vol \mathcal{B}^J(R_0) < \vol Q\,,
\end{equation}
and thus computing their quotient is not useful for counting.  From
inequality (\ref{eq:10}) we have
\begin{equation}
  \label{eq:11}
  \frac{Jq^2}{R_0^2} > 2\pi e > 17
\end{equation}
with $q=\sqrt[J]{\vol Q}$.  Other strange thing happen for large
$J$.  Let's assume for simplicity that all charges are equal to $q$.
The corner of the cell centered at the origin is located at a distance
$\frac{q}{2}\sqrt{J}$ from it, and it reaches (and surpasses) the
radius $R_0$ when
\begin{equation}
  \label{eq:12}
  \frac{Jq^2}{R_0^2} > 4\,,
\end{equation}
so we can expect the angular region near the corner to be devoid of
states.  Thus, the distribution of low $\Lambda$ states is not
isotropic in flux space, and in particular, it has no spherical
symmetry.

All these conditions coincide: when the parameter $h=\frac{Jq^2}{R^2}$
is large we cannot count by dividing volumes\footnote{With the numbers
  given above, $h=0.75$, so we can trust eq.~(\ref{eq:7}).}.  But
there are instances of the BP Landscape in which the parameter $h$ can
be large and nevertheless the model contains a huge amount of states.
In such cases the formula (\ref{eq:7}) should not be used, and another
formula is needed.  Other counting methods in the BP Landscape have
been proposed so far \cite{Shenker,BY,SV,RHM,Jul}, but all of them
have a limited range of validity.

The remainder of the paper is organized as follows.  In section
\ref{sec:omega} we will propose an exact counting formula which is
reduced to (\ref{eq:7}) for small $h$.  We will also provide an
asymptotic formula for the regime of large $h$.  In section
\ref{sec:app} we extend the method used previously to the study of
other properties of the BP Landscape, in particular the counting of
low-lying states, an estimate of the minimum value of the cosmological
constant and the possible influence of the non-trivial fraction of
nonvanishing fluxes in the KKLT moduli stabilization mechanism.  The
conclusions are summarized in section \ref{sec:conc}.

\section{The BP Landscape degeneracy}
\label{sec:omega}

In this section, we will obtain an exact integral representation for
the number of nodes of the lattice inside a sphere of arbitrary
radius, and we will analyze its main asymptotic regimes.

\subsection{The exact representation}
\label{sec:exact}

We start with the number of nodes in the lattice inside a sphere in
flux space of radius $r$.  This magnitude is called $\Omega_J(r)$
above:
\begin{equation}
  \label{eq:13}
  \Omega_J(r) = \bigl|\bigl\{\lambda\in\mathcal{L}\colon\|\lambda\|\le
  r\bigr\}\bigr|\,.
\end{equation}
In the previous equation, vertical bars denote cardinality.  An
alternative expression can be given in terms of the characteristic
function of an interval $I$
\begin{equation}
  \label{eq:14}
  \chi_I(t) =
  \begin{cases}
    1 & \text{if $t\in I$,}\\
    0 & \text{if $t\notin I$,}
  \end{cases}
\end{equation}
so that
\begin{equation}
  \label{eq:15}
  \Omega_J(r) = \sum_{\lambda\in\mathcal{L}} \chi_{[0,r]}(\|\lambda\|)
  \,.
\end{equation}
Expression (\ref{eq:15}) is exact, and the sum is extended to the full
lattice, whitout any problem given that $\chi$ function adds 1 for
each node inside the sphere, and therefore the result is always
finite.  Clearly, (\ref{eq:15}) is equivalent to directly counting
the nodes (the ``brute-force'' counting method), hence it cannot be
used in order to obtain numbers as in (\ref{eq:7}).

The density of states associated to (\ref{eq:15}) is
\begin{equation}
  \label{eq:16}
  \omega_J(r) = \frac{\partial \Omega_J(r)}{\partial r}\,.
\end{equation}
which will be called the ``BP Landscape degeneracy''.  By writing the
characteristic function in terms of the Heaviside step function
\begin{equation}
  \label{eq:17}
  \chi_{[0,r]}(\|\lambda\|) = \theta(\|\lambda\|)
   - \theta\bigl(\|\lambda\|^2 - r^2\bigr)
   \,,
\end{equation}
we obtain
\begin{equation}
  \label{eq:18}
  \omega_J(r) = 2r\sum_{\lambda\in\mathcal{L}}
  \delta\bigl(r^2-\|\lambda\|^2\bigr)
  \,.
\end{equation}
The counting function $\Omega_J(r)$ is a stepwise monotonically
non-decreasing function, and thus its derivative $\omega_J(r)$ is a
sum of Dirac deltas.  It is supported at those values of $r$ which
correspond to the values that are actually attained by the norms of
the lattice nodes.  Let $\mathcal{M}$ be the set of these values; we
have
\begin{equation}
  \label{eq:19}
  \omega_J(r) = 2r\sum_{\mu\in\mathcal{M}}\varpi_J(\mu)\delta(r^2-\mu^2)
\end{equation}
where we have defined the true degeneracy $\varpi_J(\mu)$ as the
integer-valued function which counts the number of nodes in the
lattice $\mathcal{L}$ whose norm is $\mu$, that is, the number of
decompositions of a number $\mu^2$ as a sum $\sum^J_{j=1}q_j^2n_J^2$,
for $n_1,\cdots,n_J$ integers and $q_1,\cdots,q_J$ arbitrary real
numbers.

We can express the Dirac delta which appears in (\ref{eq:18}) as a
contour integral:
\begin{equation}
  \label{eq:20}
  \delta\bigl(r^2-\|\lambda\|^2\bigr) =
  \frac{1}{2\pi i}
  \int_{\gamma} e^{s(r^2 - \|\lambda\|^2)} \dif s
  \,,
\end{equation}
where the contour is a vertical line crossing the positive real axis,
\begin{equation}
  \label{eq:21}
  \gamma=\{c+i\tau:\tau\in\R, c>0\}\,.
\end{equation}
Substituting (\ref{eq:20}) in (\ref{eq:18}), we obtain
\begin{equation}
  \label{eq:22}
  \omega_J(r) = \frac{2r}{2\pi i}
  \int_\gamma e^{sr^2}
  \Biggl[\sum_{\lambda\in\mathcal{L}}
  e^{-s\|\lambda\|^2}\Biggr]\dif s
  \,.
\end{equation}
This particular representation allows us to perform the sum extended
to the whole lattice:
\begin{equation}
  \label{eq:23}
  \begin{split}
    \omega_J(r) 
    &= \frac{2r}{2\pi i}
    \int_\gamma  e^{sr^2}
    \Biggl[\sum_{n_1\in\Z} \cdots
    \sum_{n_J\in\Z}\prod_{j=1}^J
    e^{-sq_j^2n_j^2}\Biggr]\dif s \\
    &= \frac{2r}{2\pi i}
    \int_\gamma  e^{sr^2}
    \Biggl[\prod_{j=1}^J
    \sum_{n_j\in\Z}
    e^{-sq_j^2n_j^2}\Biggr] \dif s \\
    &= \frac{2r}{2\pi i}
    \int_\gamma  e^{sr^2}
    \Biggl[\prod_{j=1}^J
    \vartheta(sq_j^2)\Biggr] \dif s
    \,.
  \end{split}
\end{equation}
The sum is hidden in the function
\begin{equation}
  \label{eq:24}
  \vartheta(s) = \sum_{n\in\Z} e^{-sn^2}\equiv \theta_3(0;e^{-s})\,,
\end{equation}
valid for $\mathrm{Re}\,s>0$, which is a particular case of a Jacobi
theta function:
\begin{equation}
  \label{eq:25}
  \theta_3(z;q) = \sum_{n\in\Z} q^{n^2}e^{2\pi i n z}\,,
\end{equation}
for complex $z$ and $q$ with $|q|<1$\footnote{The second argument $q$
  of Jacobi theta functions, the so-called \emph{nome}, shouldn't be
  mistaken with the charge $q$.}.  It satisfies the functional
equation
\begin{equation}
  \label{eq:26}
  \vartheta(s) = \sum_{n\in\Z}e^{-sn^2} =
  \sqrt{\frac{\pi}{s}}\sum_{m\in\Z}
  e^{-\frac{\pi^2 m^2}{s}}
  = \sqrt{\frac{\pi}{s}}\,\vartheta\Bigl(\frac{\pi^2}{s}\Bigr)\,,
\end{equation}
which is a consequence of the Poisson summation formula.

Our exact formula for the BP Landscape degeneracy is then
\begin{equation}
  \label{eq:27}
  \omega_J(r) = \frac{2r}{2\pi i}
  \int_\gamma  e^{sr^2}
  \Biggl[\prod_{j=1}^J
  \vartheta(sq_j^2)\Biggr] \dif s
  \,,
\end{equation}
which is an inverse Laplace transform, that is,
\begin{equation}
  \label{eq:28}
  \int^\infty_0 e^{-sr^2}\omega_J(r)\dif r = \prod_{j=1}^J
  \vartheta(sq_j^2)\,.
\end{equation}
The integration of (\ref{eq:27}) with the initial condition
$\Omega_J(0)=1$ gives
\begin{equation}
  \label{eq:29}
  \Omega_J(r) = 1 +
  \frac{1}{2\pi i}
  \int_\gamma  \frac{e^{sr^2}-1}{s}
  \Biggl[\prod_{j=1}^J
  \vartheta(sq_j^2)\Biggr] \dif s
  \,.
\end{equation}

We will close this subsection with a final remark.  By Laplace
transforming (\ref{eq:19}) and comparing it with its alternative form
(\ref{eq:28}) we obtain
\begin{equation}
  \label{eq:30}
  \sum_{\mu\in\mathcal{M}}\varpi_J(\mu)e^{-s\mu^2}
  = \prod_{j=1}^J\vartheta(sq_j^2)\,.
\end{equation}
Substituting all $q_j=1$, we can see that the possible values of the
numbers $\mu^2$ when $\mu\in\mathcal{M}$ are those which can be
represented as the sum of $J$ integer squares.  These are all the
non-negative integers, obtaining in this case
\begin{equation}
  \label{eq:31}
  1 + 2\sum_{n=0}^\infty \varpi_J(n)e^{-sn}
  = \vartheta(s)^J\,.
\end{equation}
Formula (\ref{eq:31}) is the generating function of the number of
different decompositions of a positive integer $n$ as the sum of $J$
integer squares\footnote{In number theory, the number of
  decompositions of a positive integer $n$ as the sum of $J$ squares
  is called $r_J(n)$ and its generating function is usually written
  using a variable $x=e^{-s}$ with $|x|<1$.}.  Thus, (\ref{eq:30}) can
be taken as a generalization of (\ref{eq:31}).

\subsection{The large distance (or BP) regime}
\label{sec:BP-regime}

Now we will turn to the approximate evaluations of integrals
(\ref{eq:27}) and (\ref{eq:29}).  For this purpose we need the
asymptotic behavior of $\vartheta$ function.

\begin{figure}[htbp]
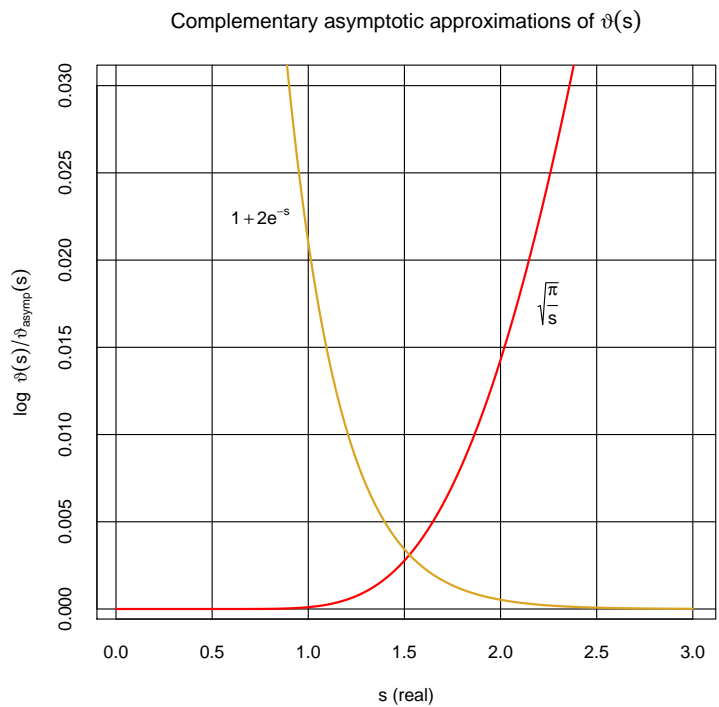

  \centering
  \image{width=10cm}{\figI}
  \caption{Complementary asymptotic regimes of the $\vartheta$
    function.  Logarithms of the quotient of $\vartheta$ and its
    asymptotes are shown.  A good approximation can be seen as a flat
    line at zero height.  Both regimes cross near $1.5$, which is the
    center of the interval $[1,2]$ where the accuracy is lower.  Note
    that the horizontal line at 0.01 signals that the quotient is 1\%
    different from 1.  All plots in this paper were done using R
    \cite{R-proj}.}
  \label{fig:theta-asymp}
\end{figure}

Function $\vartheta(s)$ has two simple asymptotic regimes for real
and positive $s$, as can be seen from the functional equation
(\ref{eq:26}):
\begin{equation}
  \label{eq:32}
  \vartheta(s)\xrightarrow{s\to 0} \sqrt{\frac{\pi}{s}}
  \quad\text{and}\quad
  \vartheta(s)\xrightarrow{s\to\infty} 1+2e^{-s}\,.
\end{equation}
We can visualize these asymptotes by plotting the logarithm of the
quotient between $\vartheta$ and each of them.  This is done in figure
\ref{fig:theta-asymp}, where we can see that the limit $s\to0$ is
(reasonably) valid for $s<1$ and the limit $s\to\infty$ is valid for
$s>2$.  In the middle regime $s\in[1,2]$ none of the two former cases
is accurate enough, and we will have a mixed, interpolating regime
between them.

The first case we will consider is $s\to0$.  In this regime, we simply
make the integration contour $\gamma$ pass near the origin in the
complex plane, where $\vartheta$ has a singularity.  Assuming that the
main contribution to the integral will come from this region, we can
replace $\vartheta$ by its asymptotic value when $s\to0$ and write
\begin{equation}
  \label{eq:33}
  \omega_J(r) \approx \frac{2r}{2\pi i}\int_\gamma e^{sr^2}
  \Biggl[\prod_{j=1}^J \sqrt{\frac{\pi}{q_j^2 s}} \Biggr] \dif s
  \,.
\end{equation}
This integral is an elementary inverse Laplace transform:
\begin{equation}
  \label{eq:34}
  \omega_J(r)  \approx
  \frac{\pi^{\frac{J}{2}}}{\vol Q}\,\frac{2r}{2\pi i} \int_\gamma
  e^{sr^2}\frac{\dif s}{s^{\frac{J}{2}}} 
  = \frac{2\pi^{\frac{J}{2}}}{\Gamma(\frac{J}{2})}\,
  \frac{r^{J-1}}{\vol Q}
  \,.
\end{equation}
Equation \eqref{eq:34} is the derivative of (\ref{eq:7}), that is, BP
count.  It is valid for large $r$ distances, because it has been
obtained using small $s$ region (note that a well known property of
Laplace transform pairs is that the asymptotic behaviour of a signal
for large $r$ is determined by the small $s$ behaviour of its
transform and vice versa).  For this reason we call the formula
(\ref{eq:34}) the \emph{large distance} regime, or BP regime.

But we can make the restriction imposed on the distance more
quantitative in the following way.  The best choice for the real part
of the contour is the saddle point of integral~(\ref{eq:34}), which is
the stationary point of the function
\begin{equation}
  \label{eq:35}
  \phi(s) = sr^2 - \frac{J}{2}\,\log s\,.
\end{equation}
This saddle point is
\begin{equation}
  \label{eq:36}
  s^* = \frac{J}{2r^2}\,,
\end{equation}
and in its vicinity we find the most important contribution to the
integral.  But the replacement of the $\vartheta(s)$ functions by its
small-$s$ behavior is valid only if the argument $s$ is less than 1
(see figure \ref{fig:theta-asymp}), so we must have
\begin{equation}
  \label{eq:37}
  s^*q_j^2 = \frac{Jq_j^2}{2r^2} < 1
  \quad\text{for all $j$}\,.
\end{equation}
Nevertheless, condition (\ref{eq:37}) does not guarantee the validity
of the replacement $\vartheta(s)\to\sqrt{\frac{\pi}{s}}$ on the
integration contour away from the real axis.  A stronger restriction
is imposed by demanding the applicability of the saddle point method
in this regime.  If the steepest descent approximation is valid, then
the main contribution to the integral comes from the vicinity of the
saddle point, which justifies the replacement of the asymptote.  The
exact and approximate evaluations of (\ref{eq:34}) will have the same
validity if both methods reach the same result.  But the saddle point
approximation on (\ref{eq:34}) has the effect of using Stirling's
approximation on the gamma function.  Thus, both methods agree if $J$
is large enough.

The correctness of the saddle point approximation of (\ref{eq:34}) can
be assessed by rewriting it in the form
\begin{equation}
  \label{eq:38}
  \omega_J(r)  \approx
  \pi^{\frac{J}{2}}\,\frac{2r}{2\pi i} \int_\gamma
  e^{sr^2 - \frac{J}{2}\log(q^2s)}\dif s
  \quad\text{with }\log q = \frac{1}{J}\sum^J_{i=1}\log q_i\,.
\end{equation}
The change of variable $q^2s = w$ transforms the exponent $\phi$ of
the integrand into
\begin{equation}
  \label{eq:39}
  \phi(w) = J\Bigl[\frac{w}{h} - \frac{1}{2}\log w\Bigr]\,,
\end{equation}
where $h=\frac{Jq^2}{r^2}$.  The validity condition of the saddle
point approximation is $\phi(w^*)\gg1$ with $w^*=h/2$ the stationary
point of $\phi$.  This condition is fulfilled if $J$ is large and
\begin{equation}
  \label{eq:40}
  \frac{w^*}{h} - \frac{1}{2}\log w^* =
  \frac{1}{2}\Bigl(1-\log\frac{h}{2}\Bigr) > 1
  \quad\Rightarrow\quad
  h < \frac{2}{e} \approx 0.736
  \,,
\end{equation}
which is analogous to the condition stated by Bousso and Polchinski
for the validity of their formula.  We have derived it as a validity
condition for the small-$s$ asymptotic regime of the exact counting
formula\footnote{Incidentally, the adimensional parameter
  $h=\frac{Jq^2}{r^2}$ occurring in \eqref{eq:39} resembles the
  t'Hooft coupling in the so-called \emph{planar limit} of field
  theory, in which the number $N$ characterizing the gauge group tends
  to infinity and the Yang-Mills coupling constant $g_{\text{YM}}$
  vanishes with the product $Ng_{\text{YM}}^2$ (the t'Hooft coupling)
  held fixed.}.

Finally, the large $J$ condition controls the validity of Stirling's
approximation for the gamma function.  But this restriction is not
needed because the integral has been done in closed form.  Thus, only
condition (\ref{eq:40}) remains.

\subsection{The small distance regime}
\label{sec:small-r}

In this case we are in the regime in which the asymptotic expansion of
$\vartheta$ for large values of its argument is valid.  We can write
$\Omega_J(r)$ in \eqref{eq:29} as
\begin{equation}
  \label{eq:41}
  \Omega_J(r) = 1 + \frac{1}{2\pi i}\int_\gamma
  f(s)e^{\phi(s)}  
  \dif s
  \,,\quad\text{with}\quad
  \begin{cases}
    f(s) = \frac{1-e^{-sr^2}}{s}\,,\\
    \phi(s) = sr^2 + \sum^J_{i=1}\log\bigl(1+2e^{-q_i^2s}\bigr)\,.
  \end{cases}
\end{equation}
The saddle-point approximation to this integral is given by
\begin{equation}
  \label{eq:42}
  \Omega_J(r) \approx 1 + \frac{1}{2\pi i}\, if(s^*)e^{\phi(s^*)}
  \sqrt{\frac{2\pi}{\phi''(s^*)}}
  \,,
\end{equation}
where $s^*$ is the stationary point of $\phi(s)$.  The saddle point
$s^*$ is a minimum for real $s$; hence, the steepest descent contour
crosses vertically the real axis and coincides locally with $\gamma$.
Unfortunately, we cannot solve the saddle point equation in closed
form for arbitrary charges.  Nevertheless, in the simplest case in
which all charges are equal $q_1=\cdots=q_J=q$, we obtain
\begin{equation}
  \label{eq:43}
  r^2 = \frac{2Jq^2}{e^{s^*q^2} + 2}
  \quad\Rightarrow\quad
  s^*q^2 = \log2\Bigl(\frac{Jq^2}{r^2} - 1\Bigr)\,.
\end{equation}
The saddle point computed through \eqref{eq:43} is consistent with the
regime of large argument of $\vartheta$ if
\begin{equation}
  \label{eq:44}
  s^*q^2 = \log2\Bigl(\frac{Jq^2}{r^2} - 1\Bigr) > 2
  \quad\Rightarrow\quad
  \frac{Jq^2}{r^2} > 1 + \frac{e^2}{2} \approx 4.694
  \,.
\end{equation}
This condition is satisfied for fixed charge $q$ and dimension $J$ if
the distance is small enough; for this reason this regime is called
the \emph{small distance} regime.

\begin{figure}[htbp]
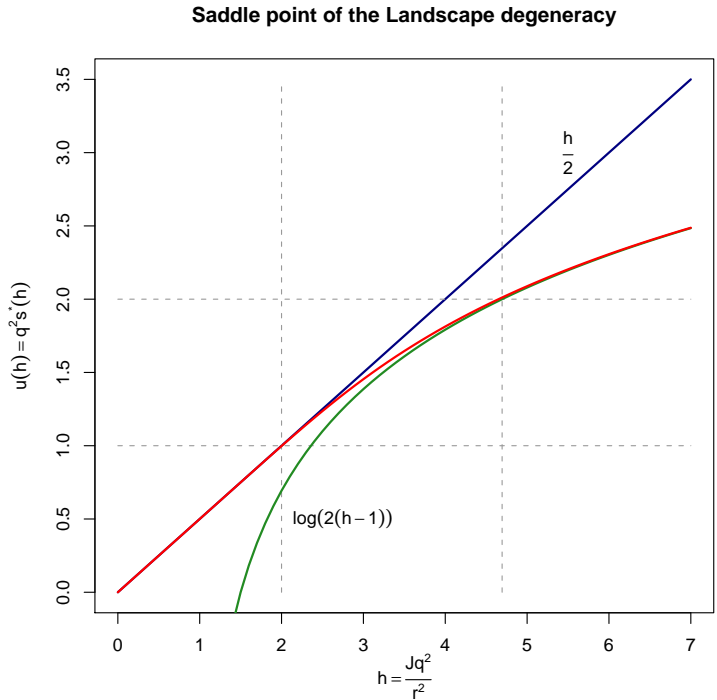

  \centering
  \image{width=10cm}{\figII}
  \caption{Numerical saddle point $u(h)$ of the $\omega_J(r)$
    integrand for equal charges.  For $h<2$ it agrees with its low-$h$
    asymptote, and for $h>5$ it agrees with its high-$h$ asymptote.
    The mixed regime, where the numerical solution smoothly
    interpolates between the asymptotes, is shown in the central
    rectangle.}
  \label{fig:saddle-point}
\end{figure}

In terms of the parameter $h=\frac{Jq^2}{r^2}$, the approximate saddle
point (which will be called $u(h)=q^2s^*(h)$) is, gluing together
eqs.~(\ref{eq:37},~\ref{eq:43})
\begin{equation}
  \label{eq:45}
  u(h) = q^2 s^*(h) =
  \begin{cases}
    \frac{1}{2}h & \text{if $h < 2$,}\\
    \log 2(h-1)  & \text{if $h > 5$.}
  \end{cases}
\end{equation}
Eq.~(\ref{eq:45}) is plotted in figure \ref{fig:saddle-point}, along
with the numerical solution obtained in a range of $h$ which is not
covered so far, but will be considered in the next subsection.

Substituting the large $h$ regime for $u(h)$ leads to
\begin{equation}
  \label{eq:46}
  \begin{split}
    \phi(s^*) &= \frac{r^2}{q^2}\,u(h) +
    J\log\bigl(1+2e^{-u(h)}\bigr)
    = J\Bigl[\frac{\log2(h-1)}{h} + \log\Bigl(\frac{h}{h-1}\Bigr)\Bigr]\,,\\
    \phi''(s^*) &= \frac{2Jq^4e^{-u(h)}}{(1+2e^{-u(h)})^2}
    = Jq^4\Bigl(\frac{h-1}{h}\Bigr)
    \,,\\
    f(s^*) &= q^2\frac{1-e^{-\frac{Ju(h)}{h}}}{u(h)}
    = q^2\frac{1-e^{-J\frac{\log2(h-1)}{h}}}{\log2(h-1)}
    \,,
  \end{split}
\end{equation}
and the saddle-point approximations for $\Omega_J(r)$ and
$\omega_J(r)$ result in
\begin{subequations}
  \begin{align}
    \Omega_J(r) &= 1 + \frac{1}{\sqrt{2\pi J}}\,
    \frac{(2h-2)^{\frac{J}{h}} - 1}{\log(2h-2)}\,
    \Bigl(\frac{h}{h-1}\Bigr)^{J+\frac{1}{2}}\,, \label{eq:47a} \\
    \omega_J(r) &= \frac{(2h-2)^{\frac{J}{h}}}{q\sqrt{2\pi h}}
    \Bigl(\frac{h}{h-1}\Bigr)^{J+\frac{1}{2}}\,. \label{eq:47b}
  \end{align}
\end{subequations}
We must note that these magnitudes depend on $r$ through
$h=\frac{Jq^2}{r^2}$.  It should be stressed that (\ref{eq:47b}) has
been obtained by using the saddle point approximation of (\ref{eq:27})
and not by differentiating (\ref{eq:47a}); although these two results
are asymptotically equivalent, they differ in subleading terms.

We will now consider the validity of formulae (\ref{eq:47a},
\ref{eq:47b}).  The saddle-point approximation is good if
$\phi(s^*)\gg 1$.  We could rewrite $\phi(s^*)$ as follows:
\begin{equation}
  \label{eq:48}
  \phi(s^* ) = J\psi(h)
  \quad\text{with}\quad
  \psi(h) = \frac{\log2(h-1)}{h} + \log\Bigl(\frac{h}{h-1}\Bigr)
  \,,
\end{equation}
and, for large $J$, demand $\psi(h)>1$ as we did before.  But this
would contradict (\ref{eq:44}), so we better choose some $J_0>2$ and
write
\begin{equation}
  \label{eq:49}
  \phi(s^*) = \frac{J}{J_0}J_0\psi(h)
  \,,
\end{equation}
which satisfies $\phi(s^*)\gg1$ if $J/J_0$ is large and
$J_0\psi(h)>1$.  For example, we have
\begin{equation}
  \label{eq:50}
  J_0 = \{1,2,3,4,5\}
  \quad\Rightarrow\quad
  h < \{2.2, 7.2, 12.5, 18.3, 24.3, 30.6\}
  \,.
\end{equation}
This restriction implies that the distance cannot be too small in
order to preserve the validity of the saddle-point approximation.

For very small $r$, we can choose a large real part of $s$ and
approximate $(1+2e^{-q^2s})^J\approx 1+2Je^{-q^2s}$, and then
$\Omega_J(r)$ is reduced to
\begin{equation}
  \label{eq:51}
  \Omega_J(r)\xrightarrow{r\to0} \theta(r^2) + 2J\theta(r^2-q^2)\,,
\end{equation}
that is, only the node at the origin and its $2J$ neighbors contribute
to $\Omega_J$.  The formula (\ref{eq:47a}) cannot reproduce this
result, and therefore there must exist a validity condition which
forbids too small distances.  This is exemplified in (\ref{eq:50}).
However, that restriction turns out to be of no importance because
close-to-the-origin nodes represent a negligible fraction of the
whole.

\subsection{The middle distance regime}
\label{sec:middle-r}

When $h$ takes a value in which no accurate asymptotic approximation
of $\vartheta$ is available, we are in the \emph{middle distance} or
\emph{crossing} regime.  In this situation the saddle point can be
computed only numerically.  This has been done in figure
\ref{fig:saddle-point} by solving the following equation
\begin{equation}
  \label{eq:52}
  -\frac{\vartheta'(w)}{\vartheta(w)} = \frac{1}{h}\,,
\end{equation}
whose solution $u(h)$ is the saddle point.  This solution can always
be obtained, with no condition on the value of $h$, but it coincides
with (\ref{eq:45}) in the specified ranges.  Using this solution, the
saddle-point approximation of $\omega_J(r)$ can be computed, but only
for large $J$, and only for not too small distances.

\begin{figure}[htbp]
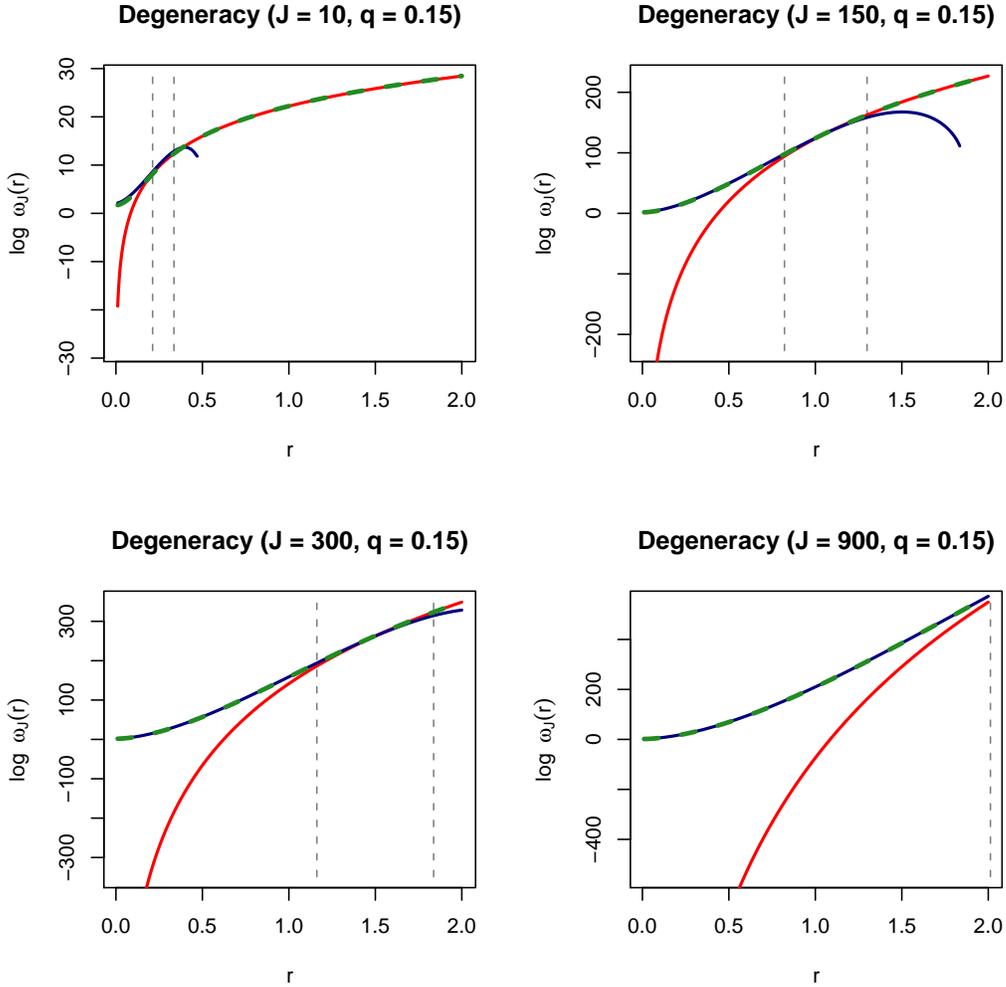

  \centering
  \image{width=14cm}{\figIII}
  \caption{Asymptotic regimes of the Landscape degeneracy
    $\omega_J(r)$ for equal charges $q=0.15$ and different $J$ as
    functions of $r$.  The actual $\omega_J(r)$ is close to the upper
    envelope of both curves (green dashed line).  For small $J$ the
    large distance or BP regime dominates (red line), but for large
    $J$ the small distance regime (blue line) spans the whole $[0,2]$
    interval.  Vertical dashed lines delimit the crossing regime,
    corresponding to $h\in[2,5]$, which is above $r=2$ in the $J=900$
    panel.}
  \label{fig:omega-J}
\end{figure}

In figure \ref{fig:omega-J}, we compare equations (\ref{eq:34}) and
(\ref{eq:47b}), displaying between them the crossing regime.  Both
small and large distance regimes show up for constant $J$ and $q$ at
different distances $r$.  Note that small $h$ corresponds to large $r$
and vice versa, but ``small'' and ``large'' distances are
$J$-dependent concepts, so that both regimes can have their own range
of validity.  Moreover, for sufficiently high $J$ almost all relevant
distances in flux space can be considered ``small''.  In such cases,
the BP formula (\ref{eq:7},\ref{eq:34}) should be replaced by the
correct asymptotic one given in (\ref{eq:47a},\ref{eq:47b}).

In the case of different charges in the small distance regime, the
approximate saddle-point equation cannot be exactly solved.  Thus, one
needs to solve the complete saddle-point equation
\begin{equation}
  \label{eq:53}
  r^2 +
  \sum^J_{j=1}q_j^2\,\frac{\vartheta'(q_j^2s)}{\vartheta(q_j^2s)} = 0
  \,.
\end{equation}
The $h$ parameter does not appear in this equation, and it is not
clear what kind of average charge must be used to define it.

However, there is a class of models in which the crossing regime
dominates over the small and large distance regimes.  It is enough to
consider a charge distribution in which the smallest and biggest
charges are well separated.  In such cases, there will be regions in
the $s$ plane where every $\vartheta$ factor could in principle lie in
a different asymptotic regime, and thus we will obtain a plethora of
intermediate regimes in which neither BP nor small distance regimes
will be accurate enough.  For large $J$, the degeneracy density
$\omega_J(r)$ can be obtained by computing the numerical solution of
(\ref{eq:52}) and then using it in the saddle point approximation of
the exact integral (\ref{eq:27}).

\section{Applications}
\label{sec:app}

In this section, we will show how $\omega_J(r)$ helps to estimate
other properties of BP models.  We will consider the number of states
in the anthropic window, the distribution of non-vanishing fluxes of a
typical state, the minimum cosmological constant and a possible
consequence on the KKLT moduli stabilization mechanism.

\subsection{Number of states in the Weinberg Window}
\label{sec:NWW}

The number of states of positive cosmological constant bounded by a
small value $\Lambda_\varepsilon$ is the number of nodes of the
lattice in flux space whose distance to the origin lies in the
interval $[R_0,R_{\varepsilon}]$, where $R_0 = \sqrt{2|\Lambda_0|}$
and $R_\varepsilon = \sqrt{2(\Lambda_\varepsilon - \Lambda_0)} \approx
R_0 + \frac{\Lambda_\varepsilon}{R_0}$ so that the width of the shell
is $\varepsilon = \frac{\Lambda_\varepsilon}{R_0}$:
\begin{equation}
  \label{eq:54}
  \mathcal{N}_\varepsilon = \Omega_J(R_\varepsilon) - \Omega_J(R_0) \approx
  \omega_J(R_0)\frac{\Lambda_\varepsilon}{R_0}\,.
\end{equation}
We remind the reader that $\Omega_J(r)$ is the number of states inside
a sphere of radius $r$ in flux space and
$\omega_J(r)=\frac{\partial\Omega_J(r)}{\partial r}$.  If
$\Lambda_\varepsilon$ is the width of the anthropic range
$\Lambda_{\text{WW}}$ (the so-called Weinberg Window), then the number
of states in it is
\begin{equation}
  \label{eq:55}
  \mathcal{N}_{\text{WW}} = \frac{\omega_J(R_0)}{R_0}\Lambda_{\text{WW}}\,.
\end{equation}
Computation of $\omega_J(R_0)$ should be done along the lines of the
previous section.  Thus, the expression \eqref{eq:55} can be used for
all values of $h$ using the relevant approximation of the exact
formula \eqref{eq:27}, which includes the BP regime as well as the
small distance and crossing regimes.

\subsection{Typical number of non-vanishing fluxes }
\label{sec:alpha-star}

As before, consider the nodes of the lattice $\mathcal{L}$ having
cosmological constant between 0 and $\Lambda_\varepsilon$.  They will
lie inside a thin shell of width $\varepsilon=R_\varepsilon-R$ above
radius $R$ (former radius $R_0$ will be called $R$ in this section).
The set of nodes inside the shell will be called $\Sigma_\varepsilon$:
\begin{equation}
  \label{eq:56}
  \Sigma_\varepsilon = \bigl\{\lambda\in\mathcal{L}\colon R\le\|\lambda\|\le
  R_\varepsilon\bigr\}
  \quad\text\quad
  \bigl|\Sigma_\varepsilon\bigr| = \mathcal{N}_\varepsilon
  \,.
\end{equation}
We will assume that $\varepsilon$ is smaller than the charges $q_i$ so
that (\ref{eq:54}) is valid but $\mathcal{N}_\varepsilon\gg1$.

Taking $J=2$, we will find at most four nodes in the shell with one
vanishing component.  Thus, the remaining states will have two nonzero
components.  In the $J=3$ case, the states in the shell are located at
the axes (at most six) with only one nonzero component, at the
coordinate planes with two nonzero components (a larger
charge-dependent number) and at the ``bulk'' of the sphere with all
three nonzero components (the most abundant).  In this way, we find
that the typical number of non-vanishing components is $J$ for the
cases $J=2,3$.

Thus, after drawing a node of the shell at random (assuming that all
nodes have the same chances of being selected), the probability of all
fluxes being different from zero will be very high.  In this section
we wonder whether it happens for all $J$.

We will answer this question by computing the fraction of states in
the shell having a fixed number $j$ of non-vanishing components.  If
all states are equiprobable, the quotient between this number and the
total number of nodes in the shell will yield the probability
distribution of the values $j$ taking into account only abundances of
states.

We can expect this probability distribution to have a peak at certain
value $j^*$.  This $j^*$ will be taken as the typical number of
non-vanishing fluxes of the states in the shell.  For small values of
$J$ we know that $j^*=J$.  We will see that this is not true for
sufficiently high $J$.

We will now outline the calculation and give the results.  The details
can be found in appendix \ref{sec:alpha-star-detail}.

For any state $\lambda\in\Sigma_\varepsilon$ having exactly $j$
non-vanishing components, we define $\alpha=\frac{j}{J}$.  When
$\lambda$ is selected at random from $\Sigma_\varepsilon$ with uniform
probability, $\alpha$ becomes a discrete random variable taking values
in the $[0,1]$ interval whose probability distribution is given by
\begin{equation}
  \label{eq:57}
  P(\alpha) =
  \frac{\mathcal{N}_\varepsilon(j)}{\mathcal{N}_\varepsilon}
  \,,
\end{equation}
where $\mathcal{N}_\varepsilon(j)$ is the number of nodes in the shell
$\Sigma_\varepsilon$ having exactly $j$ non-vanishing components.  The
formula (\ref{eq:57}) takes into account only the abundances of states
in the shell, and hence we are assuming that all states in
$\Sigma_\varepsilon$ are equally probable.

Computation of the quantity $\mathcal{N}_\varepsilon(j)$ can be
achieved using the principle of inclusion-exclusion.  For simplicity,
we will assume equal charges $q_1=\cdots=q_J=q$.  In the general
expression \eqref{eq:84} we substitute the number of nodes in the
shell (\ref{eq:54}) and the exact density of states (\ref{eq:27}),
obtaining \eqref{eq:89}.  After normalization, it results in the
following exact representation for the probability distribution (see
\eqref{eq:90} in appendix \ref{sec:alpha-star-detail}):
\begin{equation}
  \label{eq:58}
  P(\alpha) = \frac{2R}{\omega_J(R)}\binom{J}{\alpha J}
  \frac{1}{2\pi i}\int_\gamma e^{\phi(s,\alpha)}\dif s
  \quad\text{with}\quad
  \phi(s,\alpha) = sR^2 + \alpha J\log\bigl[\vartheta(q^2s) - 1\bigr]
  \,.
\end{equation}
With the assumptions made, we find that $P(\alpha)$ depends on the
radius of the $\Lambda=0$ sphere but it is independent of
$\varepsilon$.  The same method can be used for analyzing the
distribution $P(\alpha)$ over the whole Landscape, that is, inside the
sphere of radius $R_1$.  The resulting expression and the subsequent
analysis are analogous, and the result is quite similar; in
appendix~\ref{sec:alpha-star-BP} the calculation is carried out in the
BP regime.

Using the saddle-point method again, we can approximate the exact
formula (\ref{eq:58}) by (see \eqref{eq:97} in appendix
\ref{sec:alpha-star-detail})
\begin{equation}
  \label{eq:59}
  P(\alpha)  \propto e^{Js(\alpha)}
  \quad\text{with}\quad
  s(\alpha) = -\alpha\log\alpha - (1-\alpha)\log(1-\alpha) +
  \frac{1}{J}\phi(\upsilon,\alpha)\,,
\end{equation}
where $\upsilon=q^2s^*$, and $s^*$ is the stationary point of the
function $\phi(s,\alpha)$ defined in (\ref{eq:58}).  The saddle point
$\upsilon$ is a function of a single variable $h\alpha$ with
$h=\frac{Jq^2}{R^2}$, and it has two well-defined asymptotic regimes
and a crossing regime which requires numerical computation.  This is
plotted in appendix \ref{sec:alpha-star-detail}, figure
\ref{fig:saddle-point-P}.

\begin{figure}[htbp]
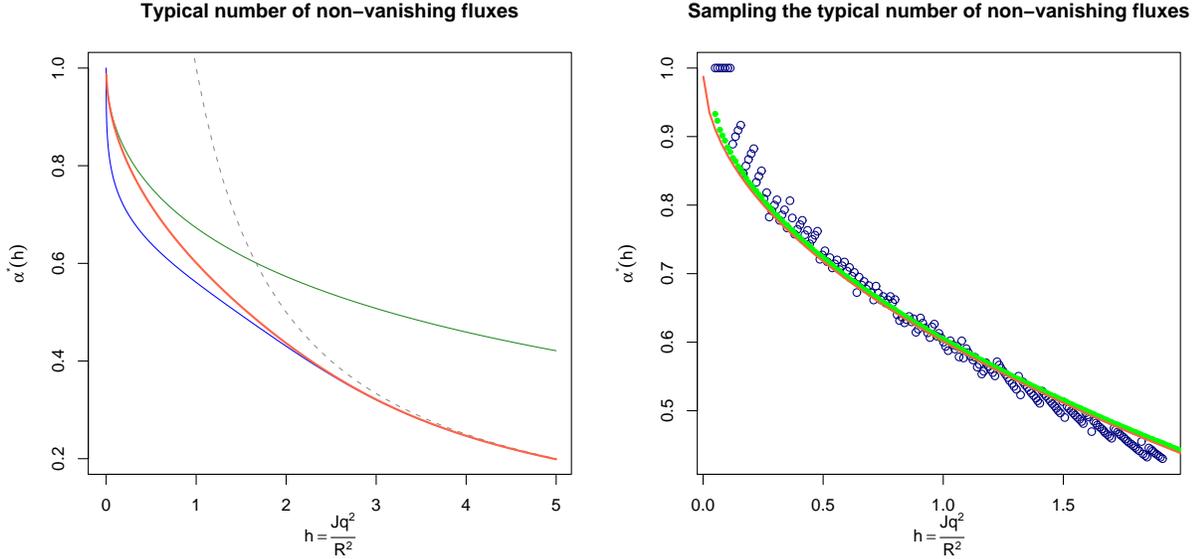

  \centering
  \image{width=\textwidth}{\figIV}
  \caption{(Left) Numerical solution $\alpha^*(h)$ along with its
    asymptotic regimes shown in equations \eqref{eq:101} (small $h$)
    and \eqref{eq:103} (large $h$).  The dashed line is the curve
    $\alpha^*=\frac{1}{h}$. (Right) Samples of the typical number of
    non-vanishing fluxes.  The two sampling methods described in the
    text have been used: The inside-shell, maximum-frequency method
    (blue hollow circles) and the inside-ball, average-frequency
    method (green bullets).  The saddle point solution displayed in
    the left panel is also shown (red line).}
  \label{fig:alpha-star}
\end{figure}

The distribution (\ref{eq:59}) has a pronounced peak located at
$\alpha^*(h)$.  This is the typical number of non-vanishing fluxes in
the shell $\Sigma_\epsilon$ (and essentially also in the whole
Landscape).  Its computation must be done numerically by solving the
following equation (see \eqref{eq:99} in appendix
\ref{sec:alpha-star-detail}):
\begin{equation}
  \label{eq:60}
  \vartheta\bigl[\upsilon(h\alpha)\bigr] = \frac{1}{1-\alpha}\,.
\end{equation}
For each positive value of $h$, \eqref{eq:60} has a unique solution
$\alpha^*(h)$ with its own regimes, which is plotted in figure
\ref{fig:alpha-star}.

Thus, $P(\alpha)$ is locally Gaussian around its peak,
\begin{equation}
  \label{eq:61}
  \log P(\alpha)\approx \log P(\alpha^*) -
  \frac{1}{2}J|s''(\alpha^*)|(\alpha-\alpha^*)^2\,, 
\end{equation}
 with standard deviation
\begin{equation}
  \label{eq:62}
  \frac{1}{\sqrt{J|s''(\alpha^*)|}} =
  \sqrt{\frac{\alpha^*(1-\alpha^*)}{J}} \le \frac{1}{2\sqrt{J}}\,.
\end{equation}
Therefore, for large $J$ the peak at $\alpha^*$ is very narrow.  We
can conclude that an overwhelming fraction of states in the shell (and
in the whole Landscape) have $J\alpha^*$ non-vanishing fluxes, and
that for high dimensions this typical number is far from $J$, which is
the typical value for the low-dimensional case considered at the
beginning of this subsection.

We should emphasize that the calculation outlined here uses the
saddle-point approximation, and therefore it is not valid for small
$J$.

Numerical searches have been carried out varying $J$ for constant $q$
and $R$, estimating $\alpha^*(h)$ by counting states.  Results are
shown in figure \ref{fig:alpha-star} versus the saddle point curve
described above.  Two sampling methods have been used:
\begin{itemize}
\item The inside-shell, maximum-frequency method samples states inside
  a shell and computes the typical number of non-vanishing fluxes as
  the value of maximum frequency.  Some advantages can be mentioned,
  such as the possibility of performing a better sampling of the true
  set we are describing, but also some disadvantages: the size of the
  sample is smaller, there is an unavoidable intrusion of exterior
  secant states in the shell \cite{RHM,Jul}, and the sample shows a
  stronger dependence with the details of the lattice.
\item The inside-ball, average-frequency method samples states inside
  a ball and computes the typical number of non-vanishing fluxes using
  the average frequency.  When compared with the previous method, this
  method has the disadvantage of sampling mostly secant states, but it
  has the advantage of accounting for bigger sample sizes.  It also
  averages over the lattice details, giving a satisfactory agreement
  with the saddle point computation.
\end{itemize}
Data used in both samples are a shell of radius $R=\sqrt{2}$ and width
$\frac{q}{2}$ with $q=0.15$, and dimensions $J$ between 2 and 200 in
the first method and between 2 and 275 in the second.  In both cases,
the $h$ parameter has been computed using averaged distances, and the
lattice details can be observed in the first sample in the form of a
jagged curve.  Note that for dimensions up to 7 the typical number of
non-vanishing fluxes is $J$ (that is, $\alpha^*=1$) but this abruptly
changes to fit the saddle point curve.

We will close this section with two remarks.  First, as the different
sampling methods considered above suggest, $\alpha^*(h)$ curve is very
robust as a property of the BP Landscape, in the sense that generic
subsets of the lattice have $\alpha^*(h)$ curves which differ only in
subleading terms.  We can see an example of this feature in appendix
\ref{sec:alpha-star-BP}.  And finally, the role played by the exact
formula \eqref{eq:27} for $\omega_J(r)$ in the computation of
$P(\alpha)$ is essential: replacing it with the BP estimate (the
``pure BP regime'') results in a probability distribution valid only
for $h < \frac{8\pi}{27}$ with the same typical fraction
$\alpha^*(h)$.  Details of this calculation are given in appendix
\ref{sec:alpha-star-BP}.

\subsection{Estimating the minimum positive cosmological constant}
\label{sec:lambda_min}

\begin{figure}[htbp]
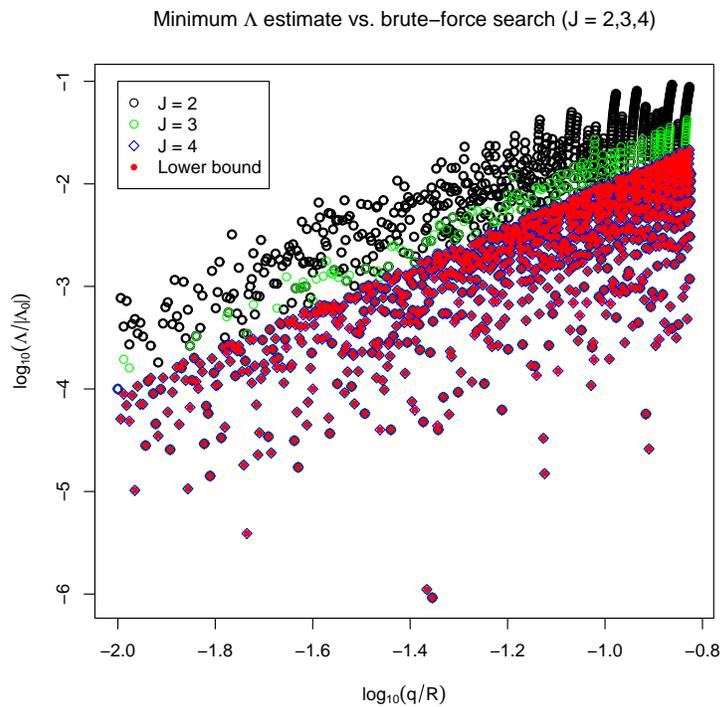

  \centering
  \image{width=10cm}{\figV}
  \caption{Actual brute-force computation of the minimum positive
    cosmological constant for equal charges in $J=2,3,4$ (hollow
    circles), along with the lower bound given in \eqref{eq:69} versus
    charge values.  The agreement is almost complete for $J=4$, and
    therefore for greater $J$ as well.}
  \label{fig:lambda-min-vs-lower-bound}
\end{figure}

In this subsection we will estimate the explicit dependence of the
minimum positive cosmological constant with respect to the parameters
of the Landscape.  We will call $\Lambda^*$ the actual minimum value,
and $\Lambda_\varepsilon$ the corresponding estimator.  We will assume
that all charges are equal, for simplicity.  In this case, we have
\begin{equation}
  \label{eq:63}
  \Lambda^* = \Lambda_0 + \frac{q^2}{2}
  \underbrace{\sum^J_{i=1}n_i^2}_{N}\,,
\end{equation}
and we should choose the smallest integer $N$ satisfying two
conditions: it should yield $\Lambda^*\ge0$, and it should be
representable as a sum of $J$ integer squares.  If we call such a
number $N_J(q)$, we have the exact formula
\begin{equation}
  \label{eq:64}
  \Lambda^* = \Lambda_0 + \frac{q^2}{2}\,N_J(q)\,.
\end{equation}
The computation of $N_J(q)$ can be avoided as we change it by another
integer $N$ satisfying only the first condition:
\begin{equation}
  \label{eq:65}
  \Lambda_0 + \frac{q^2N}{2} \ge 0
  \quad\text{but}\quad
  \Lambda_0 + \frac{q^2(N-1)}{2} < 0\,,
\end{equation}
that is,
\begin{equation}
  \label{eq:66}
  \frac{2|\Lambda_0|}{q^2} \le N < \frac{2|\Lambda_0|}{q^2} + 1\,.
\end{equation}
Thus, $N$ is the smallest integer greater than or equal to
$\frac{2|\Lambda_0|}{q^2}$, also called \emph{ceiling}:
\begin{equation}
  \label{eq:67}
  N = \biggl\lceil \frac{2|\Lambda_0|}{q^2} \biggr\rceil\,.
\end{equation}
If $N$ is decomposable as the sum of $J$ squares, we have $N_J(q)=N$,
but in general the inequality
\begin{equation}
  \label{eq:68}
  \biggl\lceil \frac{2|\Lambda_0|}{q^2} \biggr\rceil \le N_J(q)
\end{equation}
holds.  So we have a lower bound for the minimum value of the
cosmological constant, and this is our estimator:
\begin{equation}
  \label{eq:69}
  \Lambda^* \ge \Lambda_\varepsilon =
  \Lambda_0 + \frac{q^2}{2}\biggl\lceil
  \frac{2|\Lambda_0|}{q^2} \biggr\rceil\,. 
\end{equation}
It should be noted that this lower bound is independent of $J$, even
though we can expect it to work better when the error for the
replacement is small, that is, for large $J$.  In figure
\ref{fig:lambda-min-vs-lower-bound} we show the lower bound along with
the actual, brute-force computed minimum, and we find a very good
agreement for $J=4$ and greater.  In the figure we can see a straight
upper envelope of the lower bound, which can be obtained by noting
that $\lceil x \rceil < 1+x$, and thus $\Lambda_0 +
\frac{q^2}{2}\bigl\lceil \frac{2|\Lambda_0|}{q^2} \bigr\rceil <
\frac{q^2}{2}$.

Generalizing the preceding argument is not easy, because for different
charges we have
\begin{equation}
  \label{eq:70}
  \Lambda = \Lambda_0 + \frac{q^2}{2}\sum^J_{i=1}
  \frac{q_i^2}{q^2}\,n_i^2 \,,
\end{equation}
where $q$ is some average value of the charges.  The sum in
\eqref{eq:70} is not an integer, and its possible values near
$\Lambda=0$ depend strongly on the particular charge values.

We also lack a reasonable upper bound for the minimum value, because
the replacement of $N_J(q)$ by $N_J(q)+1$ in \eqref{eq:65} gives a
window of width $\frac{q^2}{2}$.  The corresponding distance in flux
space is of the order of the cell spacing, which is too big to be a
good upper bound.

We will adopt another approach from now on.  Our strategy will consist
of computing the number of states of cosmological constant between 0
and $\Lambda_\varepsilon$ and equate it to the degeneracy of the
minimum positive $\Lambda$ states computed in a different way.

Let $\lambda^*$ be a minimum cosmological constant state, that is,
$\Lambda(\lambda^*) = \Lambda^*$.  There are other valid minima of
equal value $\Lambda^*$.  Some of them can be derived from $\lambda^*$
using lattice symmetries, but there will be a set of minima which
cannot be related by any symmetries.  We will call the set of minima
modulo lattice symmetries $\Sigma^*$, and its cardinality $|\Sigma^*|$
the \emph{essential degeneracy}.  For small $J$, $\Sigma^*$ can
contain only one state, but we can expect $\Sigma^*$ to grow with $J$.

Consider a state $\lambda^*=q(n_1^*,\cdots,n_J^*)$ in $\Sigma^*$, and
let $f_k(\lambda^*)$ be the frequency of the non-negative number $k$
in the sequence $|n_1^*|,\cdots,|n_J^*|$.  Note that
$\sum_{k=0}^\infty f_k=J$, and the number of different nodes in
$\mathcal{L}$ that can be derived from $\lambda^*$ using the lattice
symmetries is
\begin{equation}
  \label{eq:71}
  \frac{J!\,2^{J-f_0}}{\prod_{k=0}^\infty f_k!}\,.
\end{equation}
The rationale behind equation (\ref{eq:71}) is to count permutations
of the $J$ components except for the components which have the same
absolute value, and multiply them by the number of different
``$J$-quadrants'' which can contain such a state (different signs of
its components), which depend on the number $J-f_0$ of non-vanishing
components.

Of course, knowing the sequence $\{f_k\}_{k=0,\cdots,\infty}$ is
equivalent to knowing the exact state $\lambda^*$, which is a very
difficult problem for large $J$.  So we can bound the true value of
the degeneracy (\ref{eq:71}) between the two extreme cases of states
having different values in all non-vanishing components (and then, all
values are non-degenerate except 0, $f_k=1$ if $k\ne0$) and states
having all non-vanishing components equal (and then, all frequencies
vanish $f_k=0$ except for two values $k\in\{0,\ell\}$ for some
$\ell$). Then, taking the same number of null components $f_0$,
\begin{equation}
  \label{eq:72}
  \frac{J!\,2^{J-f_0}}{f_0!\,(J-f_0)!} \le
  \frac{J!\,2^{J-f_0}}{\prod_{k=0}^\infty f_k!} \le
  \frac{J!\,2^{J-f_0}}{f_0!}\,.
\end{equation}
The degeneracy of the minimum $\Lambda^*$ is obtained by adding all
degeneracies of the different states $\lambda^*\in\Sigma^*$, and thus
we obtain the bound:
\begin{equation}
  \label{eq:72'}
  \sum_{\lambda^*\in\Sigma^*}
  \frac{J!\,2^{J-f_0(\lambda^*)}}{f_0(\lambda^*)!\,(J-f_0(\lambda^*))!} \le
  \sum_{\lambda^*\in\Sigma^*}
  \frac{J!\,2^{J-f_0(\lambda^*)}}{\prod_{k=0}^\infty f_k(\lambda^*)!} \le
  \sum_{\lambda^*\in\Sigma^*}
  \frac{J!\,2^{J-f_0(\lambda^*)}}{f_0(\lambda^*)!}\,.
\end{equation}
Equating the degeneracy (middle term of (\ref{eq:72'})) to the number
of states in the shell \eqref{eq:55}, we have
\begin{equation}
  \label{eq:73}
  \frac{R}{\omega_J(R)}\,
  \sum_{\lambda^*\in\Sigma^*}
  \frac{J!\,2^{J-f_0(\lambda^*)}}{f_0(\lambda^*)!\,(J-f_0(\lambda^*))!} \le
  \Lambda_\varepsilon \le
  \frac{R}{\omega_J(R)}\,
  \sum_{\lambda^*\in\Sigma^*}
  \frac{J!\,2^{J-f_0(\lambda^*)}}{f_0(\lambda^*)!}\,.
\end{equation}
Note that the shell's width is taken as $\Lambda_\varepsilon$ in
\eqref{eq:73} instead of $\Lambda^*$, because the number of states in
the shell is an estimate.  Therefore, \eqref{eq:73} can be taken as a
definition of the minimum estimator $\Lambda_\varepsilon$.

\begin{figure}[htbp]
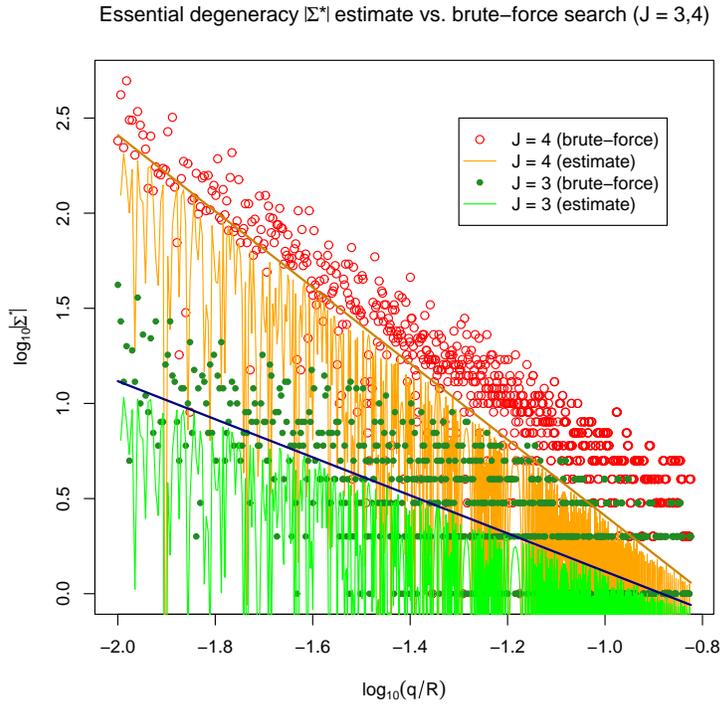

  \centering
  \image{width=10cm}{\figVI}
  \caption{Actual brute-force computation of the essential degeneracy
    $|\Sigma^*|$ of the minimum positive cosmological constant states
    for equal charges in $J=3,4$ (circles) along the lower bound given
    in \eqref{eq:77} (thin lines) versus charge values.  The strong
    oscillations of the lower bound are caused by the ceiling function
    (note logarithmic scale).  Thick lines are obtained replacing the
    lower bound of $\Lambda_\varepsilon$ by its upper envelope, and
    they represent an average asymptotic regime which is better
    followed in the $J=4$ sample.}
  \label{fig:essential-deg}
\end{figure}

So we are faced with estimating $f_0(\lambda^*)$, or equivalently, the
fraction of non-vanishing components
$\alpha(\lambda^*)=\frac{J-f_0(\lambda^*)}{J}$, in terms of which
\begin{equation}
  \label{eq:74}
  \frac{R}{\omega_J(R)}\,
  \sum_{\lambda^*\in\Sigma^*}
  \frac{J!\,2^{J\alpha(\lambda^*)}}{[J\alpha(\lambda^*)]!\,
    [J(1-\alpha(\lambda^*))]!} \le
  \Lambda_\varepsilon \le
  \frac{R}{\omega_J(R)}\,
  \sum_{\lambda^*\in\Sigma^*}
  \frac{J!\,2^{J\alpha(\lambda^*)}}{[J(1-\alpha(\lambda^*))]!}\,.
\end{equation}
The sum extended to $\Sigma^*$ can be replaced by averaging over a
probability measure of $\alpha$ restricted to $\Sigma^*$:
\begin{equation}
  \label{eq:75}
  \frac{R|\Sigma^*|}{\omega_J(R)}\,
  \int_0^1
  \frac{J!\,2^{J\alpha}}{[J\alpha]!\,
    [J(1-\alpha)]!}\dif P(\alpha|\Sigma^*) \le
  \Lambda_\varepsilon \le
  \frac{R|\Sigma^*|}{\omega_J(R)}\,
  \int_0^1
  \frac{J!\,2^{J\alpha}}{[J(1-\alpha)]!}
  \dif P(\alpha|\Sigma^*)\,.
\end{equation}
The conditional distribution $P(\alpha|\Sigma^*)$ is not known,
because the set $\Sigma^*$ is very difficult to enumerate.  We must
assume the robustness of the distribution $P(\alpha)$ computed in the
previous subsection, and approximate $P(\alpha|\Sigma^*)\approx
P(\alpha)$.  We then use the Gaussian nature of $P(\alpha)$ for large
$J$, and we perform the average simply by taking the most probable
value:
\begin{equation}
  \label{eq:76}
  \frac{R|\Sigma^*|}{\omega_J(R)}\,
  \frac{J!\,2^{J\alpha^*}}{[J\alpha^*]!\,
    [J(1-\alpha^*)]!} \le
  \Lambda_\varepsilon \le
  \frac{R|\Sigma^*|}{\omega_J(R)}\,
  \frac{J!\,2^{J\alpha^*}}{[J(1-\alpha^*)]!}
  \,.
\end{equation}
This equation is not useful because we lack an estimate for
$|\Sigma^*|$.  Nevertheless, for equal charges we can replace
$\Lambda_\varepsilon$ with \eqref{eq:69} and use \eqref{eq:76} to
obtain a lower bound for the essential degeneracy $|\Sigma^*|$:
\begin{equation}
  \label{eq:77}
  |\Sigma^*| \ge 
  \frac{\omega_J(R)}{R}\,
  \frac{[J(1-\alpha^*)]!}{J!\,2^{J\alpha^*}}
  \biggl(\Lambda_0 + \frac{q^2}{2}\biggl\lceil
  \frac{2|\Lambda_0|}{q^2} \biggr\rceil\biggr)
  \,.
\end{equation}
Thus, for the special case of equal charges, we have estimates for the
minimum value of the cosmological constant and for its essential
degeneracy.  Figure \ref{fig:essential-deg} compares the estimate
given in \eqref{eq:77} with actual brute-force computations for low
$J$.

\begin{figure}[htbp]
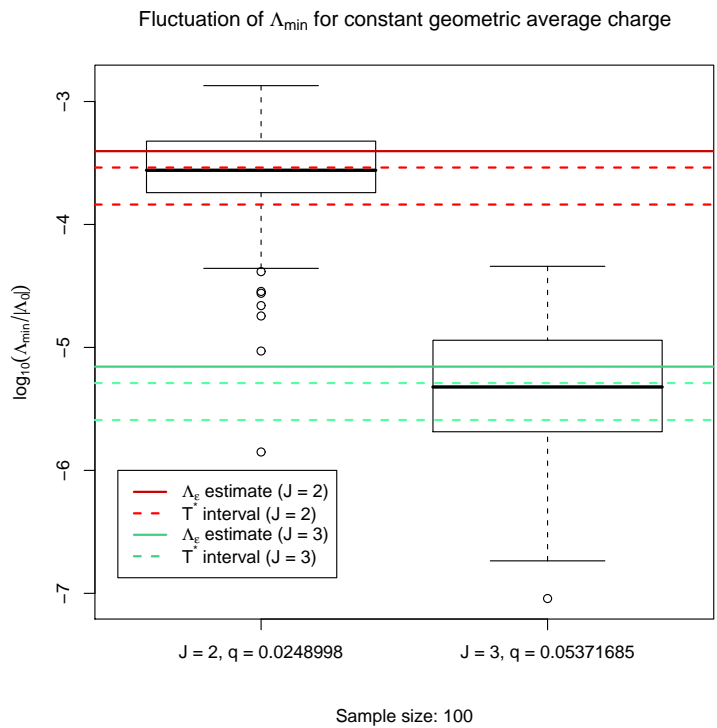

  \centering
  \image{width=10cm}{\figVII}
  \caption{Each box plot represents a sample of 100 choices of $J$
    charges with constant geometric average charge for which
    $\Lambda_{\text{min}}$ has been computed by brute-force search.
    Thick lines correspond to estimate \eqref{eq:78}, which does not
    fluctuate (at least in the BP regime).  These lines cross inside
    the boxes in good agreement with numerical data.  Dashed lines
    enclose the values provided by the ``thermodynamic'' estimator
    $T^*$, which show even better agreement.}
  \label{fig:lambda-min-different-charges}
\end{figure}

As said above, generalizing \eqref{eq:69} to the case of distinct
charges is difficult.  Nevertheless, we can assume that for charges
not only distinct but incommensurate, the essential degeneracy will be
$|\Sigma^*|=1$.  Furthermore, the symmetry degeneracy is reduced to
$2^{J\alpha}$, so that we have the estimate
\begin{equation}
  \label{eq:78}
  \Lambda_\varepsilon \approx
  \frac{2^{J\alpha^*}R}{\omega_J(R)}
  \,.
\end{equation}
Note that this formula is equivalent to the number of states in the
Weinberg Window, \eqref{eq:55}, where $\mathcal{N}_{\text{WW}}$ should
be replaced by the symmetry degeneracy and solved for the width
$\Lambda_\varepsilon$.  But we can improve this by replacing the
degeneracy $2^{J\alpha^*}$ by its mean value computed using the
Gaussian distribution $P(\alpha)$ given in formulae \eqref{eq:61} and
\eqref{eq:62}:
\begin{equation}
  \label{eq:78'}
  \langle 2^{J\alpha}\rangle_{P(\alpha)}
  = \int_0^1 2^{J\alpha}\dif P(\alpha)
  \approx \int_{\R} 2^{J\alpha}
  e^{-\frac{J(\alpha-\alpha^*)}{2\alpha^*(1-\alpha^*)}}
  \frac{\sqrt{J}\dif\alpha}{\sqrt{2\pi\alpha^*(1-\alpha^*)}}
  = 2^{J\alpha^*[1 + \frac{\log2}{2}(1-\alpha^*)]}\,.
\end{equation}
We can check this estimate with the brute-force data for low $J$ (and
thus $\alpha^*=1$) by choosing charges with constant geometric mean.
In the BP regime $\omega_J(R)$ depends only on the geometric mean and
therefore it does not fluctuate.  Figure
\ref{fig:lambda-min-different-charges} shows that \eqref{eq:78} is a
good estimator for the mean value of such fluctuations.

We can derive another estimator with an additional parameter which
allows some control of the bias by providing an interval instead of a
single value.  Nevertheless, the interval is not a confidence interval
and we lack a rigorous proof of strict inclusion.

We begin by considering the following partition function of the Landscape:
\begin{equation}
  \label{eq:125}
  Z = \sum_{\lambda\in\mathcal{L}} \theta\bigl[\Lambda(\lambda)\bigr]
  e^{-\frac{\Lambda(\lambda)}{T}}\,. 
\end{equation}
Here, the sum is carried out over all lattice nodes, but the step
function excludes all negative cosmological constant states.
$\Lambda$ plays the role of the energy, and $T$ is the associated
temperature.  Low lying states constitute the dominant contribution to
the partition function at very low temperatures, that is,
\begin{equation}
  \label{eq:126}
  Z\xrightarrow{\ T\to 0\ } e^{S^*}e^{-\frac{\Lambda^*}{T}} + e^{\hat
    S}e^{-\frac{\hat\Lambda}{T}} + \cdots\,,
\end{equation}
where $\Lambda^*$ is the actual minimum positive value of the
cosmological constant of degeneracy $e^{S^*}$, and $\hat\Lambda$ is
the first excited state of degeneracy $e^{\hat S}$.  Introducing the
gap $\delta\Lambda = \hat\Lambda - \Lambda^*$, the low temperature
free energy is
\begin{equation}
  \label{eq:127}
  F = -T\log Z \xrightarrow{\ T\to 0\ }
  \Lambda^* - TS^* - T\log\Bigl(1 + e^{\hat S -
    S^*}e^{-\frac{\delta\Lambda}{T}} + \cdots\Bigr)\,,
\end{equation}
and, at sufficiently low temperatures, namely $T\ll \delta\Lambda$,
we have the linear dependence
\begin{equation}
  \label{eq:128}
  F \approx \Lambda^* - TS^*\,.
\end{equation}
At first sight, this equation can be used to estimate the ground state
energy $\Lambda^*$ and its entropy $S^*$ by computing the free energy
in another way.  For this, we use the counting measure in the
Landscape $\Omega_J(r)$ given in \eqref{eq:29}, and write
\begin{equation}
  \label{eq:129}
  Z = \int_{R}^\infty e^{-\frac{\Lambda(r)}{T}}\dif\Omega_J(r)\,,
\end{equation}
where we have used the step function to cut off the interval of
integration.  $\Lambda$ depends on the radial variable $r$ in flux
space:
\begin{equation}
  \label{eq:130}
  \Lambda(r) = \frac{1}{2}\bigl(r^2 - R^2\bigr)\,.
\end{equation}
Note that the counting measure is reminiscent of the spherical volume
element (divided by $\vol Q$) in $J$ dimensions, and in fact reduces
to it in the BP regime, but it is really a discrete distribution whose
properties are crucially different from the continuous ones, as
will be seen below.

Replacing $\dif\Omega_J(r)=\omega_J(r)\dif r$ and using the exact
contour integral \eqref{eq:27} for $\omega_J(r)$ we obtain, upon
reversing the order of integration and changing the variable $r$ to
$u=\Lambda(r)$,
\begin{equation}
  \label{eq:131}
  \begin{split}
    Z &=
    \int_{R}^\infty e^{-\frac{\Lambda(r)}{T}}\omega_J(r)\dif r 
    = \int_{R}^\infty e^{-\frac{r^2-R^2}{2T}}
    \biggl\{
    \frac{2r}{2\pi i}\int_\gamma e^{sr^2}
    \prod_{i=1}^J \vartheta(q_i^2 s) \dif s
    \biggr\}
    \dif r \\
    &=
    \frac{1}{2\pi i}\int_\gamma
    \biggl\{
    \int_{R}^\infty e^{-\frac{r^2-R^2}{2T} + sr^2}
    2r\dif r
    \biggr\}
    \prod_{i=1}^J\vartheta(q_i^2 s) \dif s \\
    &=
    \frac{1}{2\pi i}\int_\gamma
    \biggl\{
    2 \int_{0}^\infty e^{-(\frac{1}{T} - 2s)u}
    \dif u
    \biggr\} e^{sR^2}
    \prod_{i=1}^J\vartheta(q_i^2 s) \dif s \,,
  \end{split}
\end{equation}
and the radial integral converges provided $\real\{s\} <
\frac{1}{2T}$, that is, if the pole at $\frac{1}{2T}$ is located to
the right of the integration contour $\gamma$.  Then, we have a
contour integral representation for the partition function:
\begin{equation}
  \label{eq:132}
  Z = \frac{1}{2\pi i}\int_\gamma
  \frac{e^{sR^2}}{\frac{1}{2T} - s}
  \prod_{i=1}^J\vartheta(q_i^2 s) \dif s
  \,,
\end{equation}
We can see that \eqref{eq:132} gives the initial expression
\eqref{eq:125} by expanding the $\vartheta$ sums under the product
sign.  Interchanging sum and integral, and evaluating each integral,
it turns
\begin{equation}
  \label{eq:133}
  Z = \sum_{(n_1,\cdots,n_J)\in\Z^J} \frac{1}{2\pi i} \int_\gamma
  \frac{e^{sR^2 - s\sum_{i=1}^Jn_i^2q_i^2}}{\frac{1}{2T} - s} \dif s
  \,.
\end{equation}
If the factor multiplying $s$ in the exponent is negative (thus
$\Lambda$ is positive), then we can close the contour by a large half
circle to the right; the integral on the circle vanishes, and the
contour encloses the pole at $\frac{1}{2T}$ with negative orientation,
resulting in a residue $e^{-\frac{\Lambda}{T}}$.  On the other hand,
if $\Lambda$ is negative, the contour can be closed on the left, but
the integrand has no poles in this region, and therefore the integral
vanishes.  Thus, we recover \eqref{eq:125}.

\begin{figure}[htbp]
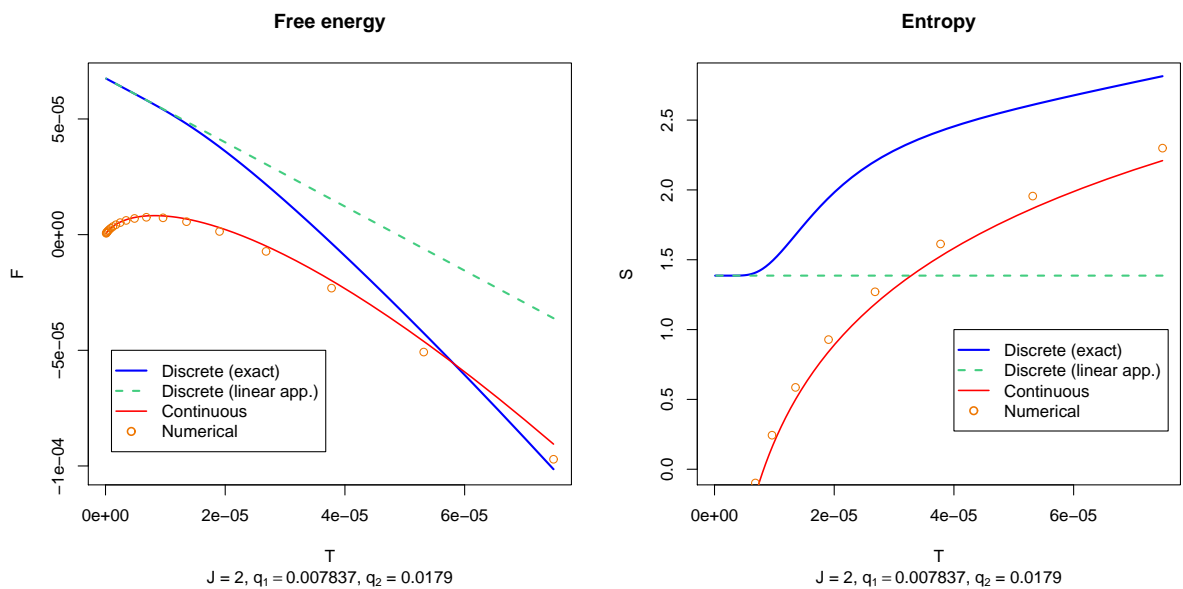

  \centering
  \image{width=\textwidth}{\figVIII}
  \caption{Comparison between thermodynamic magnitudes preserving
    discreteness or assuming continuity of the density of states.
    Near $T=0$, the exact free energy is locally linear (left panel),
    and the exact entropy reaches the value $\log(4)$ (right panel).
    Note that the numerical computation of the contour integral
    \eqref{eq:132} follows the continuous curves in both panels.  The
    exact entropy (solid thick blue line) and the continuously
    approximated one (solid thin red line) never intersect, but the
    linear approximation (green dashed horizontal line) does, defining
    the estimator $T^*$ given in \eqref{eq:139}.}
  \label{fig:discrete-vs-continuous}
\end{figure}

Equation~\eqref{eq:132} provides an independent evaluation of the
partition function by numerical computation of the contour integral.
Then we can use the values obtained in the low temperature region to
fit equation \eqref{eq:128} and estimate $\Lambda^*$ and $S^*$.
Nevertheless, the expected straight line \eqref{eq:128} is not
obtained this way, but a rather different behavior, as we now explain
(see figure \ref{fig:discrete-vs-continuous}).

Assume that we want to approximate expression \eqref{eq:129} at low
temperatures.  Changing the integration variable to $u=\Lambda(r)$
again we obtain
\begin{equation}
  \label{eq:134}
  Z = \int_{R}^\infty e^{-\frac{\Lambda(r)}{T}}\omega_J(r)\dif r
  = \int_{0}^\infty e^{-\frac{u}{T}}\omega_J
  \Bigl(\sqrt{R^2 + 2u}\Bigr) \frac{\dif u}{\sqrt{R^2 + 2u}}\,,
\end{equation}
which expresses $Z$ as the Laplace transform of $\frac{\omega_J
  \bigl(\sqrt{R^2 + 2u}\bigr)}{\sqrt{R^2 + 2u}}$.  If this
function is continuous near $u=0$, the asymptotic behavior of $Z$ is
simply
\begin{equation}
  \label{eq:135}
  Z\xrightarrow{\quad T\to0\quad} \frac{T\omega_J(R)}{R}\,, 
\end{equation}
and the thermodynamic magnitudes are, when $T\to0$,
\begin{equation}
  \label{eq:136}
  \begin{split}
    F &= -T\log\frac{T\omega_J(R)}{R} \,,\\
    S &= 1 + \log\frac{T\omega_J(R)}{R}\,,\\
    \langle\Lambda\rangle &= F+TS = T\,.
  \end{split}
\end{equation}
As seen in figure \ref{fig:discrete-vs-continuous}, the numerically
computed partition function is a good approximation for curves
computed assuming continuity, suggesting that the numerical quadrature
rule works as if the density of states were continuous, and therefore
it is not useful to estimate the ground state of the discrete
spectrum, $\Lambda^*$ and $S^*$.

It is worth emphasizing the different behavior of the thermodynamic
magnitudes between the discrete and continuous systems.  Any
computation method for $Z$ should respect the discrete nature of the
spectrum.  Otherwise the method would accurately compute the result
\eqref{eq:136}.

Nevertheless, the two behaviors coincide at high temperatures,
suggesting that there is a crossing temperature $T^*$ below which the
difference becomes drastic.  Of course, in figure
\ref{fig:discrete-vs-continuous} (right) we can see that the entropy
of the discrete system (obtained by exact computation of $Z$ using
brute-force compilation of states) and the approximate one (that of
the continuous system) never cross.  But if we approximate the former
using the approximated free energy \eqref{eq:127}, then we can find a
temperature at which the continuous and discrete entropies coincide.
This temperature $T^*$ signals the point of a drastic deviation of the
discrete system from the continuous one, and it will be interpreted as
an estimator for the ground state $\Lambda^*$.

Thus, we need the entropy in \eqref{eq:136}, which will be termed
$S_{\text{cont}}$, and the entropy derived from \eqref{eq:127}, which
is
\begin{equation}
  \label{eq:137}
  S_{\text{disc}} = S^* + \log\Bigl(1 + e^{\hat S - S^*}
  e^{-\frac{\delta\Lambda}{T}}\Bigr)
  + \frac{\delta\Lambda 
    e^{\hat S - S^*}e^{-\frac{\delta\Lambda}{T}}}
  {T\bigl(1 + e^{\hat S - S^*}
    e^{-\frac{\delta\Lambda}{T}}\bigr)}
   \,.
\end{equation}
The crossing temperature is defined by the equation
$S_{\text{disc}}=S_{\text{cont}}$, but \eqref{eq:137} contains the
degeneracy entropies $S^*$, $\hat S$ and the gap $\delta\Lambda$ as
additional parameters.  Of course, we cannot use a single equation to
fix four parameters; we will compute the degeneracies as
$2^{J\alpha^*}$ for the case of incommensurate charges (that is,
$S^*=\hat S$), and we will introduce the parameter
$\eta=\frac{\delta\Lambda}{T^*}$, so that the equation for $T^*$ reads
\begin{equation}
  \label{eq:138}
  S^* + \log\bigl(1 + e^{-\eta}\bigr)
  + \frac{\eta}{1 + e^{\eta}}
  = 1 + \log\frac{T^*\omega_J(R)}{R}
\end{equation}
and gets solved as
\begin{equation}
  \label{eq:139}
  T^* = 
  e^{S^*}\frac{R}{\omega_J(R)}
  \bigl(1 + e^{-\eta}\bigr)\,
  e^{\frac{\eta}{1 + e^{\eta}}-1}
  \,.
\end{equation}
In terms of the previous estimator $\Lambda_\varepsilon$
\eqref{eq:78}, we have
\begin{equation}
  \label{eq:140}
  T^*(\eta) = f(\eta)\Lambda_\varepsilon
  \,,\quad\text{with}\quad
  f(\eta) = \bigl(1 + e^{-\eta}\bigr)\,
  e^{\frac{\eta}{1 + e^{\eta}} - 1}
  \,,
\end{equation}
where the prefactor $f(\eta)$ is of order $\mathcal{O}(1)$ for
$0<\eta<\infty$: it is a monotonically decreasing function with
$f(0)=2e^{-1}$ and $f(\infty)=e^{-1}$.  Thus, when we let $\eta$ run
across its range, the estimator $T^*(\eta)$ spans the interval
$[1,2]e^{-1}\Lambda_\varepsilon$ instead of the single value
$\Lambda_\varepsilon$.  Nevertheless, we cannot give an argument which
favours the inclusion of the true value inside this interval.  Despite
this uncertainty, our numerical searches validate the given interval
for $J=2$ and $J=3$: it contains the median and its width is smaller
than the interquartile range (see figure
\ref{fig:lambda-min-different-charges}).

Thus, we can conclude that $T^*$ is a good estimator of the
charge-averaged minimum cosmological constant.

\subsection{A possible influence on the KKLT mechanism}
\label{sec:KKLT}

The exposition in this paper is restricted to the Bousso-Polchinski
Landscape, which is an oversimplification of the string theory
Landscape.  The Giddings-Kachru-Polchinski model \cite{GKP} is a more
realistic approach to the true Landscape, and it can be endowed with a
mechanism for fixing the moduli of the compactification manifold, the
so-called KKLT mechanism \cite{KKLT}.  In this model, the
compactification moduli are fixed by the presence of fluxes and
corrections to the superpotential coming from localized branes.  The
fixing of the moduli lead to metastable dS states by the addition of
anti-branes and exceeding flux.  The model can be further corrected to
yield inflation of the noncompact geometry by the repulsion between a
brane and an antibrane, both located at a Klevanov-Strassler throat
\cite{KKLMMT}.  The important point is that both moduli fixing and
brane inflation need the presence of flux quanta.

As far as we know, there is no combination between the BP Landscape
and the KKLT mechanism, in the sense that there is no known realistic
model in which all moduli are fixed and a large amount of three-cycles
lead to an anthropic value of the cosmological constant without any
fine-tuning at all.  Nevertheless, there are some toy models as the
six-dimensional Einstein-Maxwell theory \cite{flux-rev,6DEH} in which
it is possible to identify a Landscape with all moduli fixed, and it
has a dual model with flux; or the more sophisticated (yet unrealistic
as well) models coming from F-theory flux compactification
\cite{F-th-flux-comp} or IIB compactifications with fluxes
\cite{IIB-flux-comp}, to name a few.  Thus, it is plausible that a
complete, realistic model exhibiting all these features will be built
in the near future.

Let us assume that such a model exists.  Furthermore, let us assume
that the $\alpha^*(h)$ curve discussed in the section
\ref{sec:alpha-star} can be generalized, in the sense that, as a
characteristic feature of this conjecturally realistic model of the
string theory Landscape, there is a typical occupation number of the
fluxes that is generically different from 1 and that vanishes when the
number of fluxes $J$ is too large.  Then, we should conclude that,
generically, there will be a finite fraction $1-\alpha^*$ of
three-cycles with vanishing flux, which may be dominant if $J$ is
large and the charges $q_i$ are not too small.  This fraction of
vanishing fluxes can spoil the stabilization mechanism of the
corresponding moduli and can also spoil the brane inflation scenario
if the three-cycles located at the tip of the KS throats are devoid of
flux.

Thus, the $\alpha^*$ fraction we have found in the BP Landscape can be
an obstacle for the commonly accepted KKLT mechanism if it is also
present in a realistic Landscape model.

\section{Conclusions}
\label{sec:conc}

We have developed an exact formula for counting states in the
Bousso-Polchinski Landscape which reduces to the volume-counting one
in certain (BP) regime.  The formula might be useful in BP Landscape
examples when the parameter $h=\frac{Jq^2}{r^2}$ is large enough to
invalidate the BP count.  Applications of the formula are given which
avoid volume-counting:  counting low-lying states, computing the
typical fraction of non-vanishing fluxes or estimating the minimum
value of the cosmological constant which can be achieved in the BP
Landscape.  Numeric computations and brute-force searches have been
carried out to check the results of our analytic approximations, and
we have found remarkable agreement in all explored regimes.

In particular, we have discovered a robust property of the BP
Landscape, the curve $\alpha^*(h)$, the typical fraction of
non-vanishing fluxes, which reveals the structure of the lattice
inside a sphere for large $J$ as the union of hyperplane portions of
effective dimension near $J\alpha^*$.  This result is important in
computing degeneracies, which are used in estimating the minimum
cosmological constant.  We have not developed a formula for this
minimum, not even a probability distribution, but rather an estimator
which can produce acceptable results in predicting the mean value of
the minimum for fluctuating different charges around its geometric
mean, or essential degeneracies in the case of equal charges.

Finally, we have pointed out that if we can generalize the typical
number of non-vanishing fluxes in a realistic model describing the
string theory Landscape, then it could be an obstacle for the
implementation in the same hypotetical model of the KKTL moduli
stabilization mechanism.

Thus, the exact formula improves the way of counting as compared to
previous proposals, and should be considered in all those problems
which require state enumeration in the BP Landscape.

\section*{Acknowledgments}
\label{ack}

We would like to thank Pablo Diaz, Concha Orna and Laura Segui for
carefully reading this manuscript.  We also thank Jaume Garriga for
useful discussions and encouragement.  This work has been supported by
CICYT (grant FPA-2006-02315 and grant FPA-2009-09638) and DGIID-DGA
(grant 2007-E24/2). We thank also the support by grant A9335/07 and
A9335/10 (F\'{\i}sica de alta energ\'{\i}a: Part\'{\i}culas, cuerdas y
cosmolog\'{\i}a).

\appendix

\section{Detailed computation of the typical fraction of non-vanishing fluxes}
\label{sec:alpha-star-detail}

Consider an arbitrary subset of nodes in the lattice
$\Sigma\subset\mathcal{L}$.  We will decompose its cardinality
$|\Sigma|=\mathcal{N}_\Sigma$ as
\begin{equation}
  \label{eq:79}
  \mathcal{N}_\Sigma = \sum_{j=0}^J \mathcal{N}_\Sigma(j)\,,
\end{equation}
where $\mathcal{N}_\Sigma(j)$ is the number of states in $\Sigma$ with
exactly $j$ non-vanishing components.  If $\alpha$ is the fraction
$\frac{j}{J}$ then its probability distribution over $\Sigma$ is
\begin{equation}
  \label{eq:80}
  P_\Sigma\bigl(\alpha={\textstyle\frac{j}{J}}\bigr) =
  \frac{\mathcal{N}_\Sigma(j)}{\mathcal{N}_\Sigma}\,.
\end{equation}
This formula takes into account only abundances of states in $\Sigma$,
and hence it assumes that all states in $\Sigma$ are equally probable.

Let us introduce some notation.  We will call the set of indexes of
components $\mathcal{J}=\{1,2,\cdots,J\}$ and $L,M$ will denote any
subset of $\mathcal{J}$.  The symbol $\Sigma_L$ will denote the set of
states of $\Sigma$ having (at least) vanishing components
\emph{outside} $L$ or, in other words, the intersection between
$\Sigma$ and the subspace spanned by the directions in $L$.
$\mathcal{N}_L$ will be the number of elements of $\Sigma_L$.  Thus,
$\Sigma_\mathcal{J}$ comprises all $\Sigma$ states, that is,
$\mathcal{N}_\mathcal{J}=\mathcal{N}_\Sigma$, and $\Sigma_\emptyset$
only contains the node at the origin, so that $\mathcal{N}_\emptyset =
1$.  The inclusion-exclusion principle in Combinatorics allows us to
write
\begin{equation}
  \label{eq:81}
  \mathcal{N}_\Sigma(j) = \sum_{\substack{L\subset\mathcal{J}\\|L|=j}}
  \sum_{\ell = 1}^{j} (-1)^{j-\ell}
  \sum_{\substack{M\subset L\\|M|=\ell}} \mathcal{N}_M
  \,.
\end{equation}
The idea behind (\ref{eq:81}) is that we cannot simply sum all states
lying inside every $\Sigma_L\in\Sigma$, that is, equation
$\mathcal{N}_\Sigma(j) = \sum_{\substack{L\subset\mathcal{J}\\|L|=j}}
\mathcal{N}_L$ is not true, because of the intersections between
different subsets $\Sigma_L$.  The nodes inside these intersections 
are removed twice, and they should be added again, and so on.  This
is the cause of the alternating sign in (\ref{eq:81}).

\begin{figure}[htbp]
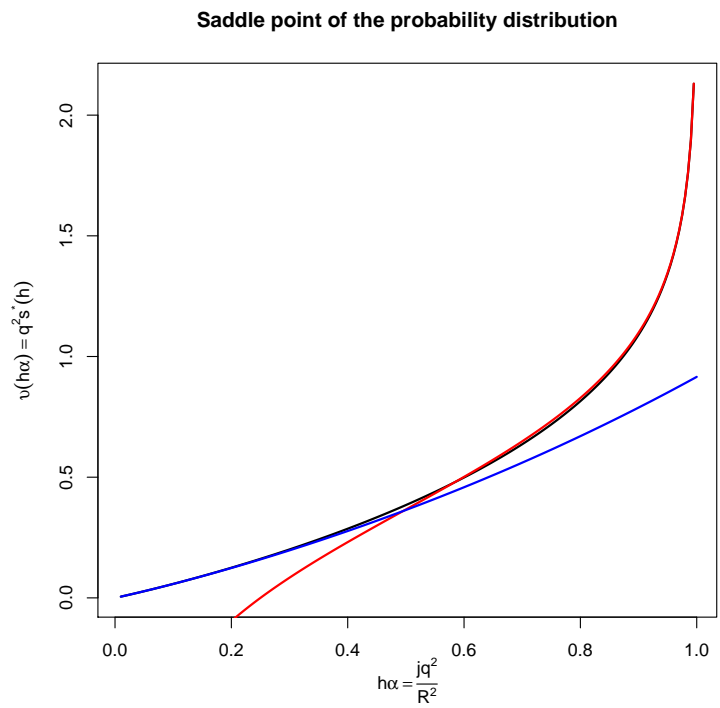

  \centering
  \image{width=10cm}{\figIX}
  \caption{Numerical saddle point $\upsilon(h\alpha)$ of the
    $P(\alpha)$ integrand for equal charges along with its two
    asymptotic regimes.}
  \label{fig:saddle-point-P}
\end{figure}

Now, we must deal with expressions of the form
\begin{equation}
  \label{eq:82}
  \sum_{\substack{L\subset\mathcal{J}\\|L|=j}}
  \sum_{\substack{M\subset L\\|M|=\ell}} \mathcal{N}_M\,.
\end{equation}
All the subsets $M\subset\mathcal{J}$ occur in the previous sum the
same number of times, which coincides with the number of supersets $L$
inside $\mathcal{J}$ (with $|L|=j$) of a given $M$ (with
$|M|=\ell<j$).  This can be computed as the number of subsets of
$\mathcal{J}\backslash M$ of exactly $j-\ell$ elements, that is,
\begin{equation}
  \label{eq:83}
  \sum_{\substack{L\subset\mathcal{J}\\|L|=j}}
  \sum_{\substack{M\subset L\\|M|=\ell}} \mathcal{N}_M
  =
  \binom{J-\ell}{j-\ell}
  \sum_{\substack{M\subset\mathcal{J}\\|M|=\ell}} \mathcal{N}_M\,.
\end{equation}
By substituting \eqref{eq:83} in \eqref{eq:81}, we get
\begin{equation}
  \label{eq:84}
  \mathcal{N}_\Sigma(j) =
  \sum^j_{\ell=1} (-1)^{j-\ell}
  \binom{J-\ell}{j-\ell}
  \sum_{\substack{M\subset\mathcal{J}\\|M|=\ell}}
  \mathcal{N}_M
  \,.
\end{equation}
We can check that equation \eqref{eq:84} satisfies the normalization
condition (\ref{eq:79}):
\begin{equation}
  \label{eq:85}
  \begin{split}
    \sum^J_{j=0}\mathcal{N}_\Sigma(j)
    &= \sum^J_{j=0} \sum^j_{\ell=1} (-1)^{j-\ell}
    \binom{J-\ell}{j-\ell}
    \sum_{\substack{M\subset\mathcal{J}\\|M|=\ell}}
    \mathcal{N}_M \\
    &= \sum^J_{\ell=0}
    \underbrace{
      \Biggl[
      \sum^J_{j=\ell} (-1)^{j-\ell}
      \binom{J-\ell}{j-\ell}
      \Biggr]
    }_{\sum^{J-\ell}_{k=0}(-1)^k\binom{J-\ell}{k} = \delta_{J,\ell}}
    \sum_{\substack{M\subset\mathcal{J}\\|M|=\ell}}
    \mathcal{N}_M \\
    &= \sum_{\substack{M\subset\mathcal{J}\\|M|=J}}
    \mathcal{N}_M = \mathcal{N}_\mathcal{J} = \mathcal{N}_\Sigma\,.
  \end{split}
\end{equation}
Now, we will take the $\Sigma$ set as the states of the lattice
$\mathcal{L}$ near the $\Lambda=0$ surface, inside a thin shell of
width $\Lambda_\varepsilon$, whose number is (\ref{eq:82})
\begin{equation}
  \label{eq:86}
  \mathcal{N}_\Sigma = \frac{\omega_J(R)}{R}\Lambda_\varepsilon\,.
\end{equation}
With different charges $q_i$, different subsets $M\in\mathcal{J}$ with
the same cardinality $|M|=\ell$ will not have the same number of
states $\mathcal{N}_M$.  However, in the simplest case where all
charges are equal, all the number of states coincide:
\begin{equation}
  \label{eq:87}
  \sum_{\substack{M\subset\mathcal{J}\\|M|=\ell}}
  \mathcal{N}_M =
  \binom{J}{\ell}\frac{\omega_\ell(R)}{R}\Lambda_\varepsilon\,,
\end{equation}
where we have used that there is $\binom{J}{\ell}$ different subsets
$M\in\mathcal{J}$ with $|M|=\ell$.  Substituting (\ref{eq:87}) in
(\ref{eq:84}) and reordering the binomial coefficients we have
\begin{equation}
  \label{eq:88}
  \begin{split}
    \mathcal{N}_\Sigma(j) &= \sum^j_{\ell=1} (-1)^{j-\ell}
    \binom{J-\ell}{j-\ell}\binom{J}{\ell}
    \frac{\omega_\ell(R)}{R}\Lambda_\varepsilon \\
    &= \frac{\Lambda_\varepsilon}{R}\binom{J}{j}
    \sum^j_{\ell=1} \binom{j}{\ell} (-1)^{j-\ell} \omega_\ell(R)
    \,.
  \end{split}
\end{equation}
Now, we can substitute the exact integral representation (\ref{eq:27})
specialized for equal charges for $\omega_\ell(R)$ and perform the
binomial sum:
\begin{equation}
  \label{eq:89}
  \begin{split}
    \mathcal{N}_\Sigma(j) &= \frac{\Lambda_\varepsilon}{R}\binom{J}{j}
    \frac{2R}{2\pi i}\int_\gamma e^{sR^2} \biggl[\sum^j_{\ell=0}
    \binom{j}{\ell}(-1)^{j-\ell}\vartheta(q^2s)^\ell \biggr]\dif s \\
    &= \frac{2\Lambda_\varepsilon}{2\pi i}\binom{J}{j}
    \int_\gamma e^{sR^2}
    \bigl[\vartheta(q^2s) - 1\bigr]^j \dif s
    \,.
  \end{split}
\end{equation}
Normalization provides the probability distribution of
$\alpha=\frac{j}{J}$ we are looking for:
\begin{equation}
  \label{eq:90}
  P(\alpha) = \frac{2R}{\omega_J(R)}\binom{J}{\alpha J}
  \frac{1}{2\pi i}\int_\gamma e^{\phi(s,\alpha)}\dif s
  \quad\text{with}\quad
  \phi(s,\alpha) = sR^2 + \alpha J\log\bigl[\vartheta(q^2s) - 1\bigr]
  \,.
\end{equation}
The next step is estimating $P(\alpha)$ using the steepest descent
method again.  The equation for the saddle point is
\begin{equation}
  \label{eq:91}
  \phi'(s) = R^2 + \alpha Jq^2\frac{\vartheta'(q^2s)}{\vartheta(q^2s) - 1}
  = 0\,.
\end{equation}
As we did before, we can find approximate expressions for the saddle
point in the two regimes of $\vartheta$ function.  If $s^*$ is the
saddle point, convenient variables are $\upsilon=q^2s^*$ and
$h=\frac{Jq^2}{R^2}$, in terms of which we have
\begin{equation}
  \label{eq:92}
  \phi(\upsilon) = \frac{R^2}{q^2}\Bigl[\upsilon +
  h\alpha\log\bigl[\theta(\upsilon)-1\bigr]\Bigr]\,.
\end{equation}
In the large $\upsilon$ regime, we have $\vartheta(\upsilon)\approx
1+2e^{-\upsilon}+2e^{-4\upsilon}$, so that
\begin{equation}
  \label{eq:93}
  \begin{split}
    \phi(\upsilon) &\approx \frac{R^2}{q^2}\Bigl[\upsilon +
    h\alpha\bigl(-\upsilon + \log2 +
    \log(1+e^{-3\upsilon})\bigr)\Bigr] \\
    &\approx
    \frac{R^2}{q^2}\Bigl[\upsilon(1-h\alpha) + h\alpha\log2 + h\alpha
    e^{-3\upsilon}\Bigr]\,,
  \end{split}
\end{equation}
and the corresponding saddle point equation has a solution
\begin{equation}
  \label{eq:94}
  \upsilon(h\alpha) =
  -\frac{1}{3}\log\frac{1}{3}\Bigl(\frac{1}{h\alpha}-1\Bigr)\,. 
\end{equation}
In the small $\upsilon$ regime, we simply have
$\vartheta(\upsilon)\approx\sqrt{\frac{\pi}{\upsilon}}$, so that
\begin{equation}
  \label{eq:95}
  \begin{split}
    \phi(\upsilon) &\approx \frac{R^2}{q^2}\Bigl[\upsilon +
    h\alpha\Bigl[\textstyle\frac{1}{2}\log\pi -
    \frac{1}{2}\log\upsilon +
    \log\Bigl(1-\sqrt{\frac{\upsilon}{\pi}}\Bigr)\Bigr]\Bigr] \\
    &\approx \frac{R^2}{q^2}\Bigl[\upsilon +
    h\alpha\Bigl[\textstyle{\frac{1}{2}\log\pi -
    \frac{1}{2}\log\upsilon}
    -\sqrt{\frac{\upsilon}{\pi}} - \frac{\upsilon}{2\pi}\Bigr]\Bigr]\,.
  \end{split}
\end{equation}
The corresponding saddle point equation is quadratic in
$\frac{1}{\sqrt{\upsilon}}$, and its solution is
\begin{equation}
  \label{eq:96}
  \upsilon(h\alpha) \approx
  \frac{4\pi}{\Bigl(\sqrt{\frac{8\pi}{h\alpha} - 3} - 1\Bigr)^2}
  \approx \frac{h\alpha}{2} +
  \sqrt{\frac{2}{\pi}}\frac{(h\alpha)^{\frac{3}{2}}}{4} +
  \frac{3(h\alpha)^2}{8\pi} +\cdots
\end{equation}
Both approximate solutions \eqref{eq:94} and \eqref{eq:96}, as well as
a numerical one, are plotted in figure \ref{fig:saddle-point-P}.
Note the logarithmic divergence at $h\alpha=1$ of the saddle point.
For larger values, $\upsilon(h\alpha)$ becomes complex, and the
integral $P(\alpha)$ begins to oscillate, losing its probabilistic
meaning.

\begin{figure}[htbp]
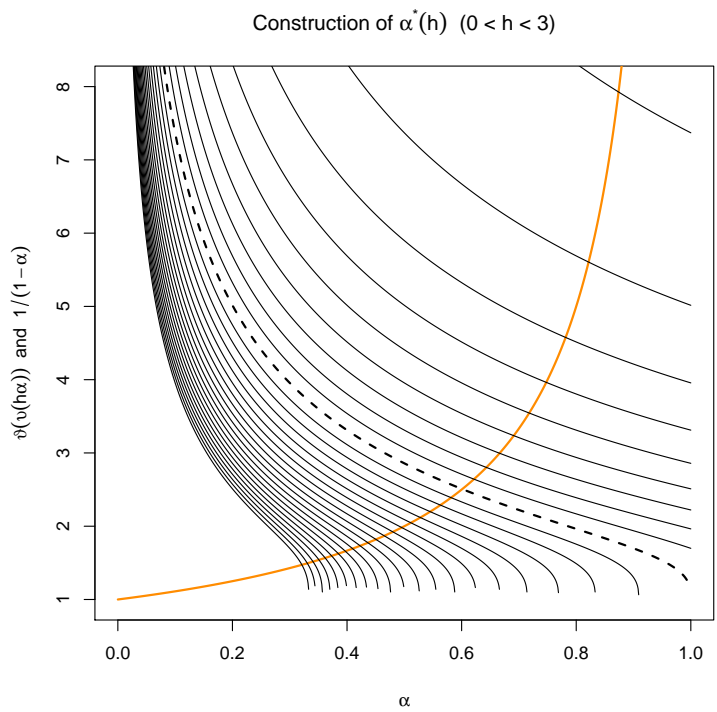

  \centering
  \image{width=10cm}{\figX}
  \caption{Construction of the solution $\alpha^*$ of equation
    (\ref{eq:99}) for $0<h<3$.  Dashed curve corresponds to $h=1$.
    For $h\ll 1$, the solution is near $\alpha=1$, and for $h\gg 1$
    the solution approaches zero.}
  \label{fig:alpha-star-build}
\end{figure}

Using the asymptotic form of the binomial coefficient, we can write
\begin{equation}
  \label{eq:97}
  P(\alpha) \propto e^{Js(\alpha)}
  \quad\text{with}\quad
  s(\alpha) = -\alpha\log\alpha - (1-\alpha)\log(1-\alpha) +
  \frac{1}{J}\phi(\upsilon,\alpha)\,,
\end{equation}
where $\phi(\upsilon,\alpha)$ also depends implicitly on $\alpha$
through $\upsilon$.  The exponent $s(\alpha)$ has a maximum
$\alpha^*$, given by the equation
\begin{equation}
  \label{eq:98}
  \begin{split}
    \frac{\dif s}{\dif \alpha} &= -\log\alpha + \log(1-\alpha) +
    \underbrace{\frac{\partial\phi}{\partial\upsilon}}_{=0}
    \frac{\dif\upsilon}{\dif\alpha} +
    \frac{\partial\phi}{\partial\alpha}\\
    &= \log\frac{1-\alpha}{\alpha} +
    \log\bigl[\vartheta\bigl(\upsilon(h\alpha)\bigr) - 1\bigr]
    = 0\,.
  \end{split}
\end{equation}
In the first equality of \eqref{eq:98} we have used the definition of
the saddle point $\upsilon$, and in the second we have used
equation~\eqref{eq:92}.  We obtain $\alpha^*$ as the unique real
solution in the $[0,1]$ interval of
\begin{equation}
  \label{eq:99}
  \vartheta\bigl[\upsilon(h\alpha)\bigr] = \frac{1}{1-\alpha}\,.
\end{equation}
Note that the saddle point $\upsilon(h\alpha)$ is defined only for
$h\alpha\le1$ (see fig.~\ref{fig:saddle-point-P}), so that the right
hand side of equation \eqref{eq:99} as a function of $\alpha$ has
domain $\alpha\in[0,\frac{1}{h}]$.  In this interval,
$\vartheta[\upsilon(h\alpha)]$ decreases from $\infty$ to 1, while the
right hand side increases from 1 to $\infty$ in $\alpha\in[0,1]$.
These conditions guarantee the existence of a unique real solution
$\alpha^*(h)\in[0,1]$ for all positive $h$.  Figure
\ref{fig:alpha-star-build} shows the construction of $\alpha^*(h)$.

For small $h$, we can use the small $h\alpha$ regime for
$\upsilon(h\alpha)$ (eq.~\eqref{eq:96}) and $\vartheta$
(eq.~\eqref{eq:32}), in which equation (\ref{eq:99}) reads
\begin{equation}
  \label{eq:100}
  \sqrt{\frac{2\pi}{h\alpha}} = \frac{1}{1-\alpha}\,,
\end{equation}
and has the solution
\begin{equation}
  \label{eq:101}
  \alpha^*(h)\approx 1 - \frac{1}{4\pi}\bigl[\sqrt{h(h+8\pi)} - h\bigr]\,.
\end{equation}
For $h\alpha$ approaching 1, we can use \eqref{eq:94} for
$\upsilon(h\alpha)$ and the corresponding regime for $\vartheta$, so
that (\ref{eq:99}) simplifies to
\begin{equation}
  \label{eq:102}
  1+2e^{\frac{1}{3}\log\frac{1}{3}\bigl(\frac{1}{h\alpha}-1\bigl)} =
  \frac{1}{1-\alpha}\,.
\end{equation}
We can isolate $h(\alpha^*)$ in (\ref{eq:102}), obtaining
\begin{equation}
  \label{eq:103}
  h(\alpha^*) \approx \frac{1}{\alpha^*\bigl(1 +
    \frac{3}{8}\,\bigl(\frac{\alpha^*}{1-\alpha^*}\bigr)^3\bigr)}\,,
\end{equation}
which for large $h$ is simply $\alpha^*\approx\frac{1}{h}$.  All these
regimes are shown in figure \ref{fig:alpha-star} along with the
numerical solution of equation (\ref{eq:99}).

\section{Typical fraction of non-vanishing fluxes in the pure BP regime}
\label{sec:alpha-star-BP}

We can repeat the computation of the probability distribution of the
fraction of non-vanishing fluxes $P(\alpha)$ (given in appendix
\ref{sec:alpha-star-detail}) in the pure BP regime step by step.
Using the previous notation, we will consider all states inside a
sphere of radius $r$ as the set $\Sigma\subset\mathcal{L}$, but we
will use the BP formula \eqref{eq:7} instead of the exact integral
representation of the number of such states (hence the name ``pure
BP''):
\begin{equation}
  \label{eq:104}
  \mathcal{N}_\Sigma = \Omega_J(r)\approx
  \frac{2\pi^{\frac{J}{2}}r^J}{J\Gamma\bigl(\frac{J}{2}\bigr)\vol Q}\,.
\end{equation}
The computation of the probability distribution is based in the
decomposition \eqref{eq:79}, which we repeat here for convenience:
\begin{equation}
  \label{eq:105}
  \mathcal{N}_\Sigma = \sum_{j=0}^J \mathcal{N}_\Sigma(j)\,,
\end{equation}
After using the inclusion-exclusion principle, we have \eqref{eq:84}
\begin{equation}
  \label{eq:106}
  \mathcal{N}_\Sigma(j) =
  \sum^j_{\ell=1} (-1)^{j-\ell}
  \binom{J-\ell}{j-\ell}
  \sum_{\substack{M\subset\mathcal{J}\\|M|=\ell}}
  \mathcal{N}_M
  \,.
\end{equation}
As before, we will assume that all charges are equal $\vol Q=q^J$.
Thus we have an equation analogous to \eqref{eq:87}
\begin{equation}
  \label{eq:107}
  \sum_{\substack{M\subset\mathcal{J}\\|M|=\ell}}
  \mathcal{N}_M
  = \binom{J}{\ell}
  \frac{\pi^{\frac{\ell}{2}}r^\ell}{\Gamma\bigl(\frac{\ell}{2}+1\bigr)q^\ell}
  = \binom{J}{\ell}
  \frac{\xi^\ell}{\Gamma\bigl(\frac{\ell}{2}+1\bigr)}
  \,,
\end{equation}
where we have used the parameter
\begin{equation}
  \label{eq:108}
  \xi = \frac{r\sqrt{\pi}}{q}\,.
\end{equation}
Plugging \eqref{eq:107} in the decomposition \eqref{eq:106},
\begin{equation}
  \label{eq:109}
  \mathcal{N}_\Sigma(j) =
  \sum^j_{\ell=1} (-1)^{j-\ell}
  \binom{J-\ell}{j-\ell}
  \binom{J}{\ell}
  \frac{\xi^\ell}{\Gamma\bigl(\frac{\ell}{2}+1\bigr)}
  = \binom{J}{j}\sum^j_{\ell=1} (-1)^{j-\ell}
  \binom{j}{\ell}
  \frac{\xi^\ell}{\Gamma\bigl(\frac{\ell}{2}+1\bigr)}
  \,.
\end{equation}
Now we take the Hankel definition of the gamma function:
\begin{equation}
  \label{eq:110}
  \frac{1}{\Gamma(x)} = \frac{1}{2\pi i}\int_C e^z z^{-x}\dif z\,,
\end{equation}
where the contour $C$ in complex $z$ plane encloses the origin coming
from $-\infty$ below the negative real axis, turning around the origin
and leaving towards $-\infty$ over the negative real axis.
By substituting
\begin{equation}
  \label{eq:111}
  \frac{1}{\Gamma\bigl(\frac{\ell}{2}+1\bigr)} = 
  \frac{1}{2\pi i}\int_C e^z z^{-(\frac{\ell}{2}+1)}\dif z
\end{equation}
in (\ref{eq:109}), we can interchange the sum and the integral and
the binomial theorem allows us to write
\begin{equation}
  \label{eq:112}
  \begin{split}
    \mathcal{N}_\Sigma(j) &= \binom{J}{j}\sum^j_{\ell=1} (-1)^{j-\ell}
    \binom{j}{\ell} \xi^\ell\frac{1}{2\pi i}\int_C e^z
    z^{-(\frac{\ell}{2}+1)}\dif z \\
    &= \binom{J}{j}\frac{1}{2\pi i}\int_C e^z\frac{\dif z}{z}
    \sum^j_{\ell=1} (-1)^{j-\ell}
    \binom{j}{\ell} \Bigl(\frac{\xi}{\sqrt{z}}\Bigr)^\ell \\
    &= \binom{J}{j}\frac{1}{2\pi i}\int_C e^z
    \biggl(\frac{\xi}{\sqrt{z}} - 1\biggr)^j
    \frac{\dif z}{z}\,.
  \end{split}
\end{equation}
By changing the $z$ variable to $s = z/\xi^2$ the contour only gets
scaled and can be deformed back to its original form, resulting in
\begin{equation}
  \label{eq:113}
  \mathcal{N}_\Sigma(j) =
  \binom{J}{j}\frac{1}{2\pi i}\int_C e^{\xi^2 s}
  \biggl(\frac{1}{\sqrt{s}} - 1\biggr)^j
  \frac{\dif s}{s}
  =
  \binom{J}{j}\frac{1}{2\pi i}\int_C e^{J\phi(s)}
  \frac{\dif s}{s}
  \,.
\end{equation}
We have defined the function
\begin{equation}
  \label{eq:114}
  \phi(s) = \frac{\pi s}{h} +
  \alpha\log\biggl(\frac{1}{\sqrt{s}} - 1\biggr)
  \,,
\end{equation}
which depends on the two parameters mainly used in the remainder of
the paper:
\begin{equation}
  \label{eq:115}
  h = \frac{Jq^2}{r^2}\qquad\text{and}\qquad\alpha=\frac{j}{J}\,.
\end{equation}
Note that $\xi=\sqrt{\frac{J\pi}{h}}$.

We can normalize $\mathcal{N}_\Sigma(j)$ dividing by
$\mathcal{N}_\Sigma$ in \eqref{eq:104}, thus obtaining the following
integral representation for the probability distribution:
\begin{equation}
  \label{eq:116}
  P(\alpha) = \Gamma\bigl({\textstyle\frac{J}{2}}+1\bigr)\xi^{-J}
  \binom{J}{j}\frac{1}{2\pi i}\int_C e^{J\phi(s)}
  \frac{\dif s}{s}
  \,.
\end{equation}
We can evaluate the integral using the steepest descent method,
assuming that $J$ is large and that $h$ and $\alpha$ are constant
parameters.  Furthermore, $h$ cannot be too large for the saddle point
approximation to remain valid.

Thus, for large $J$, $P(\alpha)$ is
\begin{equation}
  \label{eq:117}
  P(\alpha) \propto e^{Js(\alpha)}
  \qquad\text{with}\qquad
  s(\alpha) = -\alpha\log\alpha - (1-\alpha)\log(1-\alpha)
  + \phi(s^*)\,,
\end{equation}
where we have used the asymptotic expression for large $J$ of
$\binom{J}{J\alpha}$ and $s^*$ is the saddle point of $\phi(s)$ which
is compatible with the integration contour $C$.  We compute $s^*$ as a
solution of
\begin{equation}
  \label{eq:118}
  \phi'(s) = \frac{\pi}{h} - \frac{\alpha}{2}\,\frac{1}{s(1-\sqrt{s})}
  = 0\,,
\end{equation}
which can be rewritten as
\begin{equation}
  \label{eq:119}
  s\sqrt{s} - s + \nu = 0\,,
\end{equation}
with the new parameter $\nu=\frac{h\alpha}{2\pi}$.  Let us assume for
simplicity that $\nu$ is small (that is, $h$ is small); then
\eqref{eq:119} has two solutions in the positive real axis, one of
them near 0 and the other near 1.  Near $s\approx0$, $\phi''(s)\approx
\frac{\alpha}{2s^2}>0$ and near $s\approx1$,
$\phi''(s)\approx\frac{-\alpha}{4\sqrt{s}(1-\sqrt{s})^2} < 0$, so the
first corresponds to a local minimum over the positive real axis and
the second to a local maximum.  But the integration contour crosses
the positive real axis vertically from lower to upper half plane, and
therefore the integrand has a maximum along $C$ if it has a minimum
along the real axis.  So our saddle point is near 0 for small $\nu$,
and it follows from \eqref{eq:119} that we must have
\begin{equation}
  \label{eq:120}
  s^* \xrightarrow{\nu\to0} \nu\,.
\end{equation}
We can find the maximum of the ``entropy'' $s(\alpha)$ in
\eqref{eq:117}, which we will call $\alpha^*$, by solving the
stationary condition
\begin{equation}
  \label{eq:121}
  \frac{\dif s(\alpha)}{\dif\alpha} = \log\frac{1-\alpha}{\alpha} +
  \underbrace{\frac{\partial\phi(s)}{\partial s}\Bigl|_{s=s^*}}_{0}
  \frac{\dif s^*}{\dif\alpha} +
  \frac{\partial\phi(s)}{\partial\alpha}\Bigl|_{s=s^*}
  = \log\frac{1-\alpha}{\alpha} +
  \log\biggl(\frac{1}{\sqrt{s^*}} - 1\biggr)
  = 0
  \,.
\end{equation}
With the estimate \eqref{eq:120} for small $\nu$, eq.~\eqref{eq:121}
can be rewritten as
\begin{equation}
  \label{eq:122}
  \frac{1-\alpha}{\alpha} \approx \sqrt{\nu}\,.
\end{equation}
In the preceding equation, if $\nu$ is small, then its solution
$\alpha^*$ must be near 1; replacing $\alpha\approx1$ in the
denominator of \eqref{eq:122} we obtain exactly \eqref{eq:100}, which
determines $\alpha^*(h)$ for small $h$.  So this approach ends up with
the same estimate for the typical fraction of non-vanishing fluxes for
small $h$, despite the fact that we are computing this fraction
$\alpha^*$ over the whole Landscape instead of over a thin shell.
This is a strong indication that the result obtained for $\alpha^*$ is
robust, in the sense that it does not change significantly for
different generic subsets of the Landscape.

We remark that the robustness of the $\alpha^*(h)$ curve is not a
feature of spherical symmetry.  This is exemplified by the set of
secant states, which is \emph{not} spherically symmetric
\cite{RHM,Jul}.  If we repeat the computation in this appendix taking
the set of secant states at distance $r$ as the subset
$\Sigma\subset\mathcal{L}$ instead of the ball of radius $r$, we would
obtain the same estimate.  Thus, we should consider $\alpha^*(h)$ a
robust property of the BP Landscape.

The preceding computation used the BP regime \eqref{eq:104}, and we
have mentioned above that $h$ should be small for the validity of the
saddle-point approximation.  Now we can wonder, how big can $h$ be before
invalidating the approximation.  Can we somehow continue this result
to higher values of $h$?

\begin{figure}[htbp]
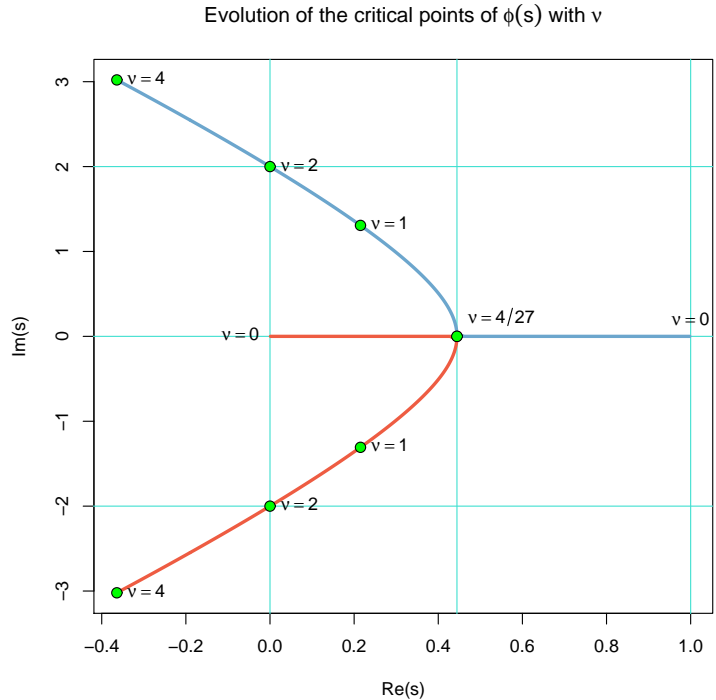

  \centering
  \image{width=10cm}{\figXI}
  \caption{Evolution in complex $s$-plane of the critical points of
    the $\phi$ function \eqref{eq:114}, exponent of the integral
    representation of $P(\alpha)$ in the pure BP regime, versus the
    $\nu$ parameter.  The saddle point $s^*$ plotted in red (which
    begins at $s=0$ and enters the lower half plane for
    $\nu>\frac{4}{27}$) is compatible with the integration contour
    $C$, which means that $C$ can be deformed to coincide with the
    steepest descent contour crossing trough $s^*$.}
  \label{fig:s-star-vs-nu}
\end{figure}

The answer to this latter question is negative, as can be seen in
figure \ref{fig:s-star-vs-nu}, which, in turn, answers also the former
question.
In figure \ref{fig:s-star-vs-nu} we show the saddle points of
$\phi(s)$ as they evolve in complex $s$-plane from $\nu=0$ to $\nu=4$.
Both of them start at $s=0,1$ respectively for $\nu=0$ and evolve
along the positive real axis until they coincide at $s=\frac{4}{9}$
for $\nu=\frac{4}{27}$.  This coincidence can be seen by writing
\eqref{eq:119} in a variable $x=\sqrt{s}$ and considering its
derivative, which is a polynomial whose roots $x^*$ are the critical
points of $p(x)$:
\begin{equation}
  \label{eq:123}
  p(x) = x^3 - x^2 + \nu \xrightarrow{\frac{\dif}{\dif x}}
  3x^2-2x = 0
  \quad\Rightarrow\quad
  x^*\in\Bigl\{0,\frac{2}{3}\Bigr\}
  \quad\Rightarrow\quad
  p(x^*) \in \Bigl\{\nu,\nu-\frac{4}{27}\Bigr\}\,,
\end{equation}
that is, $p(x)$ has two critical points with values above and below
the $x$-axis only if $\nu < \frac{4}{27}$.  For $\nu>\frac{4}{27}$
both critical points have positive values, and the roots of $p(x)$
leave the real axis.

Only the root near 0 is compatible with the integration contour $C$,
that is, $C$ can be deformed to meet the steepest descent contour
crossing by the saddle point near $s=0$.  For $\nu>\frac{4}{27}$ both
roots leave the real axis; as a consequence, the integral $P(\alpha)$
remains real but becomes non-positive and rapidly oscillating, thereby
losing its meaning as a probability distribution.  Therefore, the pure
BP regime is only valid for $\nu<\frac{4}{27}$, that is, for
\begin{equation}
  \label{eq:124}
  h < \frac{8\pi}{27} = 0.93084\,.
\end{equation}
This value represents the upper limit of validity of the pure BP
regime in the computation of $P(\alpha)$.  We can intuitively
understand the existence of this limit by considering that the
inclusion-exclusion principle \eqref{eq:106}, being an alternating
sum, is very sensitive to inaccurate computations of the cardinals of
the subsets appearing in the sum.  Therefore, when we use the BP
formula \eqref{eq:104} for computing the cardinals of the subsets with
growing $h$, that is, for decreasing $r$, the error in the formula
causes strong oscillations in $P(\alpha)$.

\end{document}